\documentclass[twocolumn,showpacs,preprintnumbers,amsmath,amssymb,prb]{revtex4-1}
\usepackage{soul}
\usepackage[bbgreekl]{mathbbol}
\usepackage{graphicx}% Include figure files
\usepackage{dcolumn}% Align table columns on decimal point
\usepackage{bm}% bold math
\usepackage{mathtools}
\usepackage{hyperref}
\usepackage{color}
\usepackage[a4paper]{anysize}
\graphicspath{{fig/}{img/}}

\newcommand{\wqd}{  {   \frac {\sqrt{\omega_T \omega_B} } 2  e^{-qd}}}
\newcommand{\sqd}{ \frac 1 2 \sqrt{\frac {\omega_B }{\omega_T}}e^{-qd}}

\newcommand{\qn}{{\bf  q}}

\newcommand{\rn}{{\bf r}}

\newcommand{\yn}{{\bf y}}
\newcommand{\zn}{{\bf z}}

\newcommand{\un}{{\bf u}}
\newcommand{\Jn}{{\bf J}}
\newcommand{\En}{{\bf E}}
\newcommand{\An}{{\bf A}}

\def\gsim{\lower.35em\hbox{$\stackrel{\textstyle>}{\textstyle\sim}$}}
\def\lsim{\lower.35em\hbox{$\stackrel{\textstyle<}{\textstyle\sim}$}}

\begin{document}
\title{Quantum Plasmons in Double Layer Systems}
\author{Luis Brey$^1$ and H.A.Fertig$^{1,2,3}$}
\affiliation{ $^1$ Instituto de Ciencia de Materiales de Madrid (CSIC), Cantoblanco, 28049 Madrid, Spain}
\affiliation{$^2$ Quantum Science and Engineering Center, Indiana University, Bloomington, IN 47408}
\affiliation{$^3$ Department of Physics, Indiana University, Bloomington, IN 47405}
\date{\today}
\begin{abstract}
Plasmons are fundamental excitations of metals which can be described in terms of electron dynamics, or in terms of the electromagnetic fields associated with them.  In this work we develop a quantum description of plasmons in a double layer structure, treating them as confined electromagnetic modes of the structure.  The structure of the resulting bosonic Hamiltonian indicates the presence of virtual plasmons of the individual layers which appear as quantum fluctuations in the ground state. For momenta smaller than the inverse separation between layers, these modes are in the ultrastrong coupling regime.  Coherence terms in the Hamiltonian indicate that modes with equal and opposite momenta are entangled.  We consider how in principle these entangled modes might be accessed, by analyzing a situation in which the conductivity of one of the two layers suddenly drops to zero.  The resulting density matrix has a large entanglement entropy at small momenta, and modes at $\pm \qn$ that are inseparable.  More practical routes to releasing and detecting entangled plasmons from this system are considered.
\end{abstract}
%\pacs{73.20.-r, 78.67.-n, 73.20.Mf, 73.21.Ac}
\maketitle
%%%%%%%%%%%%%%%%%%%%%%%%%%%%%%%%%%%%%%%%%%%%%%%%%%%%%%%%%%%%%
%  SECTION INTRODUCTION
%%%%%%%%%%%%%%%%%%%%%%%%%%%%%%%%%%%%%%%%%%%%%%%%%%%%%%%%%%%%%
\par \noindent
\section{Introduction}
\label{sec:introduction}
In metals, Coulomb interactions among carriers yield self-sustained collective charge density oscillations, which when quantized are called plasmons \cite{Pines:1952aa,Pines:1956aa,Sawada:1957aa,Vignale-book,Pitarke:2007aa}. In three dimensions, because of the long-range nature of the interaction, the plasmon spectrum is gapped, and in the long wavelength limit its energy is given by $\hbar\Omega$, where $\Omega$ is the classical plasma frequency.
%Energy loss spectroscopy experiments have measured the plasma frequency in metals and its quantization\cite{Ruthermann,Lang}.
Two-dimensional (2D) realizations of plasmons
can also be found at the surface of a metal \cite{Ritchie:1957aa}, or in 2D conducting materials \cite{Ando:1982aa,Stauber:2014aa,Stauber:2013ac,Stauber17}. In these systems
%the charge density oscillates   along the  system and
the plasmon spectrum is gapless, vanishing as $\sim\sqrt{q}$, where $q$ is the 2D momentum.
% and then for low density systems the plasmons energies are in the infrared.
% opening a variety of optoelectronics technological possibilities.
%Plasmons in quantum wells have been observed in far-infrared optical absorption\cite{Allen:1977aa} and resonant inelastic  light-scattering\cite{Pinczuk:1989aa}.
An interesting distinction between three- and two-dimensional plasmons is that in the latter, the electric field associated with the density oscillations exists outside the material, allowing strong coupling to other electromagnetic sources and modes.  Indeed in such systems, the collective modes may be described completely in terms of electromagnetic degrees of freedom, so that the plasmons may be understood as confined light modes
%Charge density oscillations  are  strongly coupled with electromagnetic fields and in 2D,
%the plasmons can be obtained from the Maxwell equations, where the conductivity of the metal sheet appears  in the boundary conditions for the electromagnetic fields.
%plasmons merge and can be understood as
and are known as
%When 2D plasmons are strongly coupled to other external photon modes, the resulting modes are known as
surface-plasmon-polaritons.\cite{nikitin-book,Nikitin:2011aa,Koppens:2011ab,Slipchenko:2013aa}
%\textit{(Luis feels the last two sentences are ambiguous, that it sounds like we are saying all plasmons are SPP's.  I am not sure I see the ambiguity.  Do we need to modify?)}

Among systems that support 2D plasmons, graphene has emerged as a particularly remarkable platform.  Graphene is a single layer of carbon atoms arranged in a honeycomb network; in pristine form it is a semimetal, but it can easily be made metallic using the electric field effect \cite{Guinea_2009,Katsnelson-book}.
It is an interesting material in the context of plasmons because it is open to the environment,
allowing their direct visualization
%from  the interference between the tip emitted and the edge scattered plasmons
in near-field microscopy experiments \cite{Chen:2012aa,Fei:2012aa,Fei:2013aa}. Moreover graphene may be patterned or gated to create plasmonic metamaterials \cite{Ju11}, and is particularly attractive for photonics and nano-optoelectronics \cite{Chen:2012aa,Bonaccorso:2010aa,Farmer:2015aa} because it supports long propagation lengths \cite{Ni15}, and can be tuned such that the relevant
frequencies are in the terahertz range.

As with light, many interesting and useful physical phenomena associated with plasmons in graphene can be understood in a purely classical framework \cite{Chen:2012aa,Fei:2012aa,Fei:2013aa,Liu:2015aa,Alcaraz18,Gopalan:2018aa,Ju11,Nikitin:2012aa,Vacacela-Gomez:2016aa,Peres:2012aa,
Sunku:2018aa,Brey:2020aa,Brey:2020ab,Stauber:2013aa}.  However, as in optics \cite{Grynberg-book,Vogel-Welsch-Book}, the underlying quantum structure of the plasmons allows for behaviors that are purely quantum mechanical in nature.
For example, a quantum treatment of surface plasmons is necessary for modeling stimulated emission of quantum emitters \cite{Torma:2015aa}, quantum correlations between plasmons \cite{Ciuti:2005aa,Berthel:2016aa,Sun:2022aa}, or coupling effects mediated by plasmons \cite{Chang:2006aa,Gonzalez-Tudela:2013aa,Gonzalez-Tudela:2011aa}.

%The quantum nature of plasmons also  plays an important role in the understanding of the electron energy loss through metal foils\cite{Marton:1962aa}.
%Also zero point energy of quantum plasmons is an important contribution to the total energy of an electron gas.\cite{Pines1964ElementaryEI}
\begin{figure}
\includegraphics[width=8.cm]{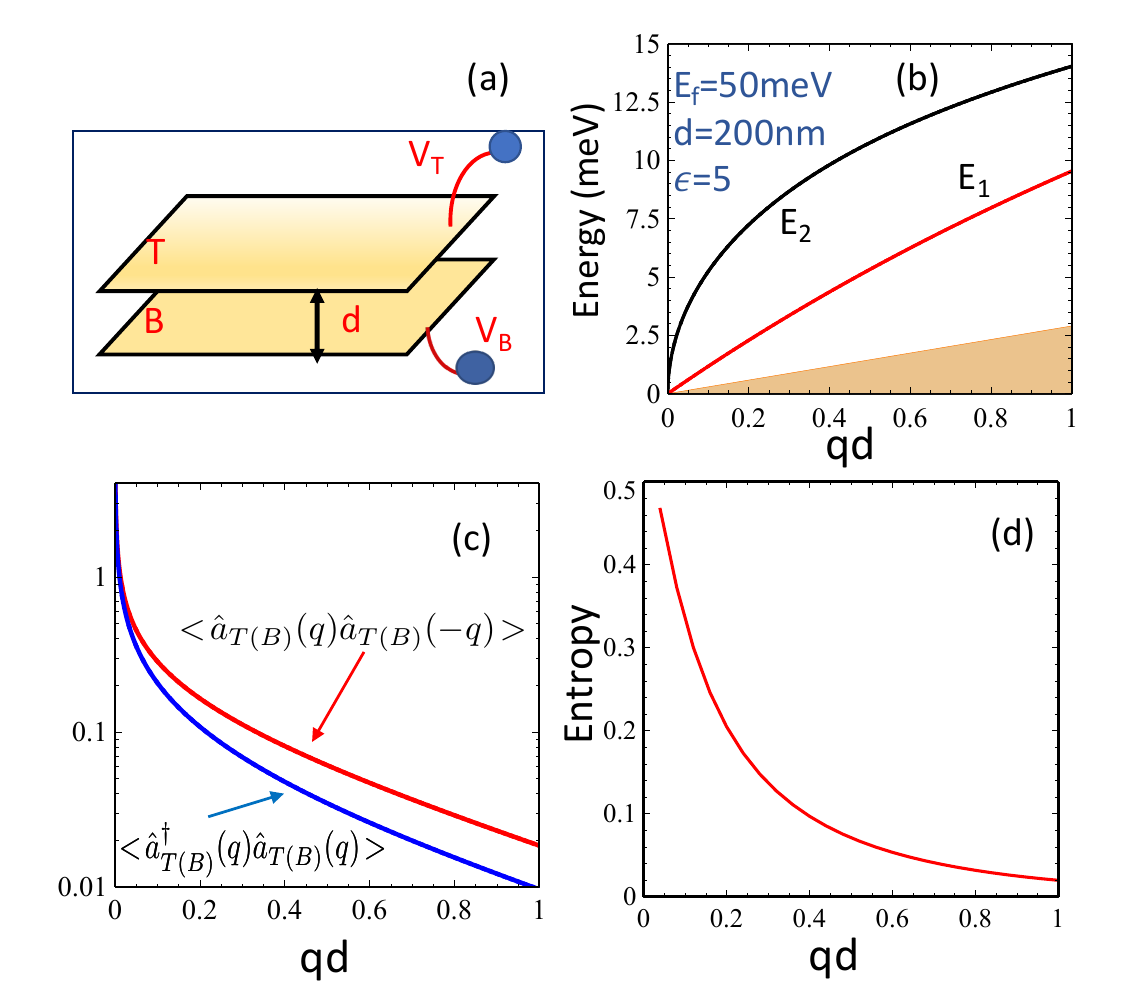}
\caption{(a) Scheme of the DL system. (b) Dispersion of the acoustic, $E_2$, and optical plasmons, $E_1$, of a DL for the parameters indicated in the figure. The plasmons do not overlap with the
electron-hole continuum, shadow region, indicating they  have  long lifetime. (c) Population and intralayer coherence of virtual  plasmons in T or B layers as function of $qd$. (d) Entropy associated with the entanglement of plasmons in T and B layers. }
\label{Figure1}
\end{figure}

In this work we investigate the quantum properties of plasmons in a double layer (DL) graphene system, as illustrated in Fig. \ref{Figure1}(a).
The layers, labeled T (top) and B (bottom), are separated by an insulator barrier thick enough to prevent electron tunneling, but thin enough
for the interlayer Coulomb interaction to be important \cite{Ponomarenko:2011aa,Stauber:2012aa,Rodrigo:2017aa}.  The qualitative behavior of collective plasmon modes in such double layer systems has been well-known for some time \cite{Das-Sarma:1981aa}, in particular that they support in-phase optical modes and out-of-phase acoustic modes (see Fig. \ref{Figure1}(b)).  We demonstrate that in a quantum description,
the
ground state contains a finite number of virtual plasmons in the T and B layers, as illustrated in Fig. \ref{Figure1}{c}. Energy conservation dictates that these virtual excitations cannot be withdrawn from the system directly \cite{Ciuti:2005aa}; however, in the presence of time-dependence in the Hamiltonian, their presence can have highly non-trivial consequences.
We consider a protocol in which the electron density in one layer is dropped suddenly to zero.  Because of the presence of the virtual plasmons, plasmons that emerge in the remaining charged layer have strong quantum correlations.  In particular, we find a large entanglement entropy among these plasmons.  Moreover, we show that the density matrices for plasmons with momenta $\pm {\bf q}$ are generically inseparable \cite{Duan_2000,Braunstein-book}, so that correlations between these two modes are intrinsically quantum in nature; i.e., they cannot be explained within any classical description.

%As a particular case, we also study  plasmons in a double layer system  in which the top layer is strongly modulated  in density and can be considered as a  periodic array  of quantum dots.  When the separation between dots is large enough  the collective excitations, plasmons, in the isolated top layer
%are gapped due to the confinement and dispersion-less because of the weak interaction between quantum dots. Then, in the ground state of the DL plasmon system, the virtual plasmons creates an
%entanglement between local excitations of the quantum dots of the top layer.

\section{
Preliminaries: 2D Plasmons as Confined Electromagnetic Modes}
\label{sec:preliminaries}
In what follows we model 2D metals as dissipationless conductors, characterized by an optical conductivity that for small momenta and frequencies takes  the form \cite{Wunsch:2006aa,Hwang:2007aa,Brey:2007aa} $\sigma (\omega; E_F)$=$ i \frac {D} {\omega}$, where $D=\frac {e ^2 E_F}{\hbar ^2 \pi}$ is the Drude weight and $E_F$ is the Fermi energy.  In  graphene $E_F$ is related to the density of carriers, $n_0$, through the relation
$E_F=\hbar v_D \sqrt{\pi n_0}$, with $v_D$ the velocity of the graphene Dirac points \cite{Guinea_2009, Katsnelson-book, Stauber:2014aa}.
Semiclassically, in 2D the plasmon frequency, $\omega _q $=$\sqrt{ \frac { D}{2 \epsilon _d \epsilon _0} q}$ (with $\epsilon _d $ is the dielectric constant of the surrounding medium), and the corresponding electric and magnetic fields can be obtained
from Maxwell's equations, with proper matching of the fields across the 2D metal sheet, taking into account the optical conductivity $\sigma (\omega; E_F)$ \cite{nikitin-book,Supp1}.

We proceed to write a quantum Hamiltonian for the electromagnetic field associated with plasmons, making some simplifying assumptions. Specifically we assume the semi-static limit, which is appropriate when
the plasmons wavelength is much smaller than the light wavelength at the same frequency.
% the plasmon velocity $\omega _q /q$ is much smaller than the speed of light.  \textit{(Do we get in trouble at small q?  The plasmon speed diverges there.)}
In this situation the magnetic field contribution to the electromagnetic energy is negligible (see SI.)
Our Hamiltonian then becomes \cite{ Elson:1971aa,Gruner:1996aa,Archambault:2010aa,Hanson:2015aa,Ferreira:2020aa,Supp1}
\begin{widetext}
\begin{equation}
\hat H   =     \frac {  \epsilon _0 \epsilon _d} 2      \int  \int d \rn dz   {\hat \En} (\rn,z) {\hat \En} (\rn,z)  +
\frac 1 2   \int  \int d \rn dz   D \delta (z)    {\hat \An} (\rn,z) {\hat \An} (\rn,z)=\sum_{\qn} \frac {\hbar \omega _q } 2 \left ( \hat a _{\qn} \hat a ^{\dagger}_{\qn} + \hat a ^{\dagger}_{\qn} \hat a _{\qn} \right ),
\label{QH}
\end{equation}
\end{widetext}
where the operator $\hat a _{\qn}$ annihilates a plasmon with 2D momentum $\qn$ and frequency $\omega _{q}$.  In terms of $\hat a _{\qn}$, $\hat a _{\qn}^{\dag}$ the electric field and vector potential operators are
\begin{eqnarray}
{\hat \En} (\rn,z)  &
=&   \sqrt{ \frac {\hbar \omega _q }{ 2 \epsilon _0 \epsilon_d S  }}   e ^{i \qn  \rn } \un (\qn,z)  \hat a _{\qn} + h.c., \nonumber \\
{\hat \An} (\rn,z) & = &  -i \sqrt{ \frac {\hbar }{ 2 \epsilon _0 \epsilon_d S \omega _q  }}   e ^{i \qn  \rn } \un (\qn,z)  \hat a _{\qn} + h.c., \nonumber \\
\label{QH1}
\end{eqnarray}
where $S$ is the sample area and the vectors $\un (\qn,z)$ are given by
\begin{equation}
{\bf u}(\qn,z)=   e^{-q |z|} \sqrt { \frac q  2 } \left (i \frac {\qn } q -  \frac {z}{|z|}
 \hat {\zn} \right )  \, .
 \label{QH2}
\end{equation}
In Eq. \ref{QH} the first term is the energy stored in the electric field energy and the second term, which is nonzero only in the conducting layer, represents the kinetic energy of the charge
carriers.

\section{
Plasmons in Double Layer System}
\label{sec:plasmons}
Consider two  metallic sheets, T and B,  located at $z=\frac d 2$ and $z=- \frac d 2$ respectively. The separation $d$ is assumed large enough that  electron tunneling between the layers can be neglected. The layers are connected to metallic contacts
that define their Fermi energies, as illustrated in Fig. \ref{Figure1}(a). We assume both layers have the same density of carriers and so the same plasmon dispersion $\omega _q$ when isolated.

In the double layer system, the total electric field is the sum of fields generated by the plasmons in the T and B layers. The total electric field operator becomes
%\begin{equation}
${\hat \En}(\rn,z) \! = \! \hat {{\bf E}}^T ( \rn,z)\!+ \! \hat {{\bf E}}^B ( \rn,z)$, with
%\end{equation}
\begin{equation}
\hat {{\En}}^{T (B)}( \rn,z) \! = \!  \sum _{\qn} \sqrt{ \frac {\hbar \omega _q}{ 2 \epsilon _0 \epsilon_d S  }} e^{i \qn \rn}  {\bf u}(\qn,z  \mp  \frac d 2  )  \, \hat a ^{T(B)} _{\qn} +h.c.,
% \nonumber \\
%\hat {{\En}}^B( \rn) & =& \sum _{\qn} \sqrt{ \frac {\hbar \omega _q}{ 2 \epsilon _0 \epsilon_d S  }} e^{i \qn \rn}  {\bf u}(\qn,\kappa,z \! +\! \frac d 2  )  \, \hat a ^B _{\qn} +h.c.
\end{equation}
where $ \hat a ^{T(B)}_{\qn}$ annihilates a plasmon with momentum $\qn$ in the T(B) layer.
Coupling between plasmons in the T and B layers appears in the cross term of the electric field energy, $ \hat V \! = \! \epsilon _0 \epsilon \int d \rn dz \,  {\hat {{\bf E}}^T ( \rn,z)} {\hat {{\bf E}}^B ( \rn,z) }$. By performing the integral of this contribution  the DL Hamiltonian becomes
\begin{equation}
\hat H  =  \hat H_T+\hat H_B+ {\hat V} \label{HDL}
\end{equation}
with
\begin{equation}
\hat H_{T(B)}   =    \sum _{\qn }  \hbar \omega _q  \left (  \hat a ^{T(B)}   _{\qn} { } ^{\dagger}  \, \hat a ^{T(B)}   _{\qn} +\frac 12    \right ) \nonumber
\end{equation}
and
%\end{eqnarray}
%\begin{eqnarray}
%{\hat V} \! & = &  \! \sum _{\qn} \frac { \hbar \sqrt{\omega _T (q)  \omega _B(q) }} 2 e^{-q d}  \nonumber \\ & \times &
\begin{equation}
{\hat V}  =   \sum _{\qn} \frac { \hbar \omega _q} 2 e^{-q d}
% \nonumber \\ & \times &
\! \left (   \hat a ^T   _{\qn} { } ^{\dagger}  \! \hat a ^B   _{\qn} \! + \!  \hat a ^B   _{\qn} { } ^{\dagger}  \! \hat a ^T   _{\qn} \! -\!  \hat a ^T   _{\qn}   \! \hat a ^B   _{-\qn} \!
-\!  \hat a ^{B   \dagger} _{\qn}   \, \hat a ^{T  \dagger}   _{-\qn}
\right ).
\nonumber
\end{equation}
%and $\hat H    _{T(B)}$=$ \sum _{\qn }  \hbar \omega _q  \left (  \hat a ^{T(B)}   _{\qn} { } ^{\dagger}  \, \hat a ^{T(B)}   _{\qn} +\frac 12    \right ) $

Eq. \ref{HDL} represents a system of  four coupled quantum harmonic oscillators
for which we could have  obtained   the normal modes before quantizing, with the result
$\omega _{1(2)}(q) =\omega _q \sqrt{ 1\mp e ^{-qd}}$. These excitations are the acoustic and optical plasmons of the DL system \cite{Das-Sarma:1981aa,Hwang:2009aa}.  The present formulation allows one to go beyond a classical treatment
% for the case of equally doped layers, i.e. $\omega_T (q) $=$\omega_B(q)$=$\omega (q)$.
to analyze the quantum properties of the coupled plasmon system.
The first two terms of  the coupling  $\hat V$ are the resonant part of the interaction, and describe the creation of a plasmon in one layer and the annihilation of a plasmon in the other, while conserving momentum. The last two terms correspond to processes which are non-conserving in the number of plasmons: they simultaneously create or annihilate pairs of plasmons
with opposite wavevectors.  In quantum optics such contributions to the Hamiltonian are known as counterrotating (CR) terms \cite{Grynberg-book,Vogel-Welsch-Book,Kavokin-Book,Torma:2015aa,Frisk-Kockum:2019aa,Kirton:2019aa}.
The coupling coefficient $\Omega_{\qn} \equiv \frac {  \omega (q)} 2 e^{-q d}$ is the $q$-dependent Rabi frequency of the DL plasmon system.

In problems involving coupling of matter and light, CR terms are often neglected.  This rotating wave approximation (RWA) works well when the coefficients of the CR terms are sufficiently small
\cite{Grynberg-book,Vogel-Welsch-Book,Kavokin-Book,Torma:2015aa,Frisk-Kockum:2019aa,Kirton:2019aa}.
For the present problem the RWA yields plasmon frequencies
$
 \omega ^{1 (2)}_{RWA} $ = $ \omega (q) \left ( 1 \mp \frac {e^{-qd} } 2 \right  )
$,
which is a good approximation to the normal mode frequencies when $qd \gg 1$, but fails significantly when $qd \lesssim 1$.
The former can be understood as representing the eigenfrequencies of the DL system to first order in perturbation theory in $e^{-qd}$.  At this level of approximation,
self-consistency in the electric fields associated with the plasmons is not fully implemented. The importance of the CR terms at long wavelengths is an indication that this system is in the ultrastrong coupling limit, defined as situations in which the Rabi frequency is not small compared to the uncoupled oscillator frequencies \cite{Torma:2015aa,Frisk-Kockum:2019aa}, in this case $\omega(q).$
%Here,
%Rabi frequency $\Omega _{\qn}/\omega(\qn) $=$e^{-qd}/2$  that for vales of $qd \sim 0.2$  can be as high as 0.4.

For long wavelength plasmons, it is necessary to diagonalize the Hamiltonian including the CR terms. Because the Hamiltonian is bilinear in the field operators, it can be diagonalized through a
Bogoliubov-Hopfield transformation \cite{Hopfield:1958aa,Ciuti:2005aa}.
This involves a symplectic transformation of the T and B creation and annihilation operators, which maintains the bosonic commutation relations of the operators while
bringing the Hamiltonian, Eq. \ref{HDL}, into diagonal form (see SI \cite{Supp1}).
Explicitly, with a transformation of the form
\begin{eqnarray}
\hat b_{1(2)} (q)&= & \frac 1 {\sqrt 2 }   {\Big (}  \cosh {\theta _{1(2)} }
(\hat a ^T (\qn) \! \mp \!  \hat a^B(\qn))  \nonumber \\  & + &  \sinh{\theta _{1(2)} }
(\hat a ^T { } ^{\dagger}(-\qn) \!  \mp \!  a^B{ } ^{\dagger}(-\qn)),
{\Big  ) }
\label{NewBasis}
\end{eqnarray}
where
$e ^{-2 \theta_i} = {\omega _i} (q) /{ \omega _q }$, the transformed Hamiltonian becomes
\begin{equation}
\hat H = \sum _{\qn, i=1,2}
 {\hbar {\omega _i}(q) }    \left ( \hat b _ i ^{\dagger} (\qn) \hat b _ i (\qn)   +\frac 1 2 \right ).
\end{equation}
The frequencies   $ {\omega _i} (q)  $ are identical to the normal mode frequencies obtained in a classical calculation \cite{Supp1}.

\section{
Virtual Plasmons and Entanglement}
\label{sec:virtual}
For finite layer separation the ground state of the system is not the vacuum $|0\rangle$ of the uncoupled DL system, for which $ \hat a^B(\qn))|0$$\rangle$=$\hat a^T(\qn))|0$$\rangle$=$0 $, but rather the vacuum $|G\rangle$ of
coupled modes, which satisfies $ \hat b _1(\qn))|G\rangle$=$\hat b  _2 (\qn))|G\rangle$=$0 $.  Because the $b_i({\bf q})$ annihilation operators are linear combinations of annihilation and creation operators  $a^{T,B}({\bf q})$ and $a^{T,B\,\dag}({\bf q})$ associated with
individual layers, the DL vacuum  contains a non-vanishing number of plasmons in the T and B layers  with   inter- and intra-layer  coherence.  Explicitly, inverting Eq. \ref{NewBasis} one finds the expectation values
\begin{eqnarray}
& \langle \hat a^{T(B)}{}^{\dagger} (\qn) \hat a^{T(B)} (\qn) \rangle= \! \frac  { \sinh ^2 \theta _2 +  \sinh ^2 \theta _1 } 2\, \, , \nonumber  \\
&\langle \hat  a^T{}^{\dagger} (\qn) \hat a^B (\qn)  \rangle  = \! \frac  {\sinh ^2 \theta _2 -  \sinh ^2 \theta _1} 2 \, \, , \nonumber \\
&\langle \hat  a^T{} (\qn) \hat a^T (-\qn) \rangle  = \! \frac  {\sinh  {2 \theta _1}+  \sinh {2 \theta _2}  } 4  \, .
\label{coherence}
\end{eqnarray}
In Fig. \ref{Figure1}(c) we plot $ \langle \hat a^{T \dagger} (\qn) \hat a^T (\qn) \rangle $ and $\langle \hat a^{T } (\qn) \hat a^T (-\qn) \rangle$ as functions of $qd$. These plasmons in the vacuum state are virtual and cannot be extracted from the isolated DL system.
Note that these expectation values are non-vanishing because of the presence of CR terms in the Hamiltonian, Eq. \ref{HDL}. Since the
Rabi frequency decreases exponentially with $qd$, the average number and the coherence of single-layer plasmons decreases rapidly with $qd$.

Because $|G\rangle$ is a state with well-defined, vanishing total in-plane momentum, the presence of
intralayer coherence between plasmons of opposite wavevector is an indication that they are entangled: they must be introduced into the ground state in equal and opposite pairs, as can be seen explicitly in the CR terms of $\hat{H}$.  We quantify the degree of entanglement as follows.
The Hamiltonian (Eq. \ref{HDL}) is expanded in the number state basis \cite{Makarov,Zhou_2020}
%of the number state representation of the plasmons in the T and B layers %$|\{\bn \}> \equiv
$|n^T _\qn, n^B _\qn, n^T _{-\qn}, n^B _{-\qn} \rangle$, where $n^{\nu} _{\qn}$ is the number of plasmons with momentum $\qn$ in  layer $\nu$, and then diagonalized, to
obtain the coefficients $ \langle  n^T _\qn, n^B _\qn, n^T _{-\qn}, n^B _{-\qn}  |G$$\rangle$.  These are the probability amplitudes which encode different numbers of single layer plasmons in the ground state.
%, the  DL plasmons vacuum can be expressed as\begin{equation}|G> = \sum _{\{\bn \}} ^{\infty} C_{\{\bn \}} |  \{ \bn \}>\end{equation}
%The coefficients $ C_{\{\bn \}}$=$<\{ \bn \}|G>$
We then compute
%To quantify the entanglement in the system we compute
%the von Neumann entropy associated with
the reduced density matrix for the subsystem of plasmons in layer T with momentum $\qn $,
\begin{eqnarray}
\rho^{(\qn)}(n^T_{\qn}; n^{' T}_{\qn})  =  \quad\quad\quad\quad\quad\quad\quad\quad\quad\quad\quad\quad\quad\quad\quad\quad&& \nonumber \\
\sum _{ n^B_{\qn}, n^B_{-\qn}, n^T_{-\qn}  } \! \! \! \!\! \! \! \! \langle n^T_{\qn},n^B_{\qn}, n^B_{-\qn} ,n^T_{-\qn} | \hat \rho | n^{'T}_{\qn},n^B_{\qn}, n^B_{-\qn} ,n^T_{-\qn} \rangle, \nonumber \\
\end{eqnarray}
%\textit{(Is there a question about the above matrix?)}
where $\hat \rho$ is the density operator for the full system. The von Neumann entropy associated with $\rho^{(\qn)}$ quantifies the entanglement between plasmons of wavevector ${\bf q}$ in the top layer and all other plasmon modes (in both the top and bottom layer) \cite{Girvin-book},
\begin{equation}
%S_{T,\qn} \! \! =\! - \rm{Tr} [ M_{ n^T_{\qn}, n^{T'}_{\qn}} \ln {M_{ n^T_{\qn}, n^{T'}_{\qn}}} ] \! = \!  - \! \sum _{i} |\lambda _i |^2 \ln{ |\lambda _i|^2 }
S_{T,\qn} \! \! = \!  - \! \sum _{i} |\lambda _i |^2 \ln{ |\lambda _i|^2 },
\end{equation}
where the sum is over the eigenvalues $\lambda_i$ of the reduced density matrix, $\rho ^{(\qn)}_{ n^T_{\qn}, n^{' T}_{\qn}}  $.
%  The reduced entropy of the four different subsystems is the same, i.e. $S_{T,\qn} $=$S_{T,-\qn} $=$S_{B,\qn} $=$S_{B,-\qn} $\cite{Girvin-book}.
In Fig. \ref{Figure1}(d) we plot the entropy $S_{T,\qn}$ as a function of the parameter $qd$. This entropy is non-vanishing because the virtual excitations in the vacuum of the DL involves quantum entanglement among plasmons in the  four subsystems ($\pm {\bf q}$ for T and B layers.) Again, note that inclusion of the CR terms is crucial to obtaining a non-vanishing entanglement entropy.

\section{
Releasing Entangled Plasmons}
\label{sec:releasing}
The populations of plasmons in the individual  layers, which are present in the ground state of the system, are virtual.  In order to access them, the quantum Hamiltonian
must be perturbed or modulating in some way \cite{Ciuti:2005aa}.
One protocol by which this could be done in principle involves a non-adiabatic time-dependent  perturbation.  In particular a sudden drop in potential, for example in the bottom layer, can deplete its charge, leaving only the top layer as a remaining host for plasmons.  (Since in our approach this involves at temporal change in boundary conditions for the electromagnetic field, this is a realization of the dynamical Casimir effect \cite{Nation_2012}.)
If this switch-off time is shorter than the inverse of a typical Rabi frequency,% \textit{(Probably need to say here that in practice this is difficult)}
to a first approximation this represents a sudden change in the Hamiltonian.  The loss of mobile charge in the bottom layer eliminates the electric field that it previously generated, so that in Eq. \ref{HDL}, $\hat{H} \rightarrow \hat{H}_T$.  Although the plasmon degrees of freedom from the bottom layer are formally present in the Hilbert space, they no longer contribute to the dynamics of the system.  The initial state of the system after the sudden change is an excited state of $\hat{H}_T$ (in the Hilbert space of both layers), and plasmons which were previously virtual become detectable.  Importantly,
the intralayer coherence between top layer plasmons with momenta $\qn$  and -$\qn$, $\langle$$  \hat  a^T{} (\qn) \hat a^T (-\qn)\rangle$, indicates they will be entangled.
Thus this geometry in principle offers
a source of counter-propagating, entangled plasmons.

In principle we would like to quantify the degree of entanglement between plasmon modes with equal and opposite momenta.  However, the state of these modes by themselves is characterized not by a wavefunction but rather by a density matrix, arrived at by tracing out all the other modes from the pure state density matrix.  In this situation one evaluates entanglement-like correlations by examining the form of the density matrix.  To do this we define the two-mode density matrix,
\begin{widetext}
\begin{equation}
\label{separable}
\rho^{(2)}(n^T_{\qn},n^T_{-\qn};n^{\prime T}_{\qn},n^{\prime T}_{-\qn}) =
\sum _{ n^B_{\qn}, n^B_{-\qn}}\langle n^T_{\qn},n^B_{\qn}, n^B_{-\qn} ,n^T_{-\qn} | \hat \rho | n^{\prime T}_{\qn},n^B_{\qn}, n^{B}_{-\qn} ,n^{\prime T}_{-\qn} \rangle.
\end{equation}
We would like to know if $\rho^{(2)}$ can be written in the form
\begin{equation}
\label{sepform}
\rho^{(2)}(n^T_{\qn},n^T_{-\qn};n^{\prime T}_{\qn},n^{\prime T}_{-\qn})=
\sum_i p_i \rho^{(\qn)}_i(n^T_{\qn};n^{\prime T}_{\qn}) \otimes \rho^{(-\qn)}_i(n^T_{-\qn};n^{\prime T}_{-\qn}),
\end{equation}
where $0 < p_i < 1$ are real numbers representing probabilities for different states, and $\rho_i^{(\qn)}$ are single $\qn$ mode density matrices for the top layer.  If Eq. \ref{sepform} holds, then $\rho^{(2)}$ represents a mixture of unentangled states.  By contrast, if $\rho^{(2)}$ {\it cannot} be written in this form, the two modes are ``inseparable'' \cite{Braunstein-book}, the generalization of entanglement to a setting where some quantum degrees of freedom have been traced out.
\end{widetext}

A test of whether Eq. \ref{sepform} holds was developed in Ref. \onlinecite{Duan_2000} and is implemented as follows.  Writing $\hat{a}_{\pm\qn}^{T} \equiv \frac{1}{\sqrt{2}} \left(\hat{x}_{\pm}+i\hat{p}_{\pm} \right)$, with $[\hat{x}_s,\hat{p}_{s^{\prime}} ] = i \delta_{s,s^{\prime}}$, $s,s^{\prime}=\pm$, one can form EPR-like quadrature operators
\begin{eqnarray}
  \hat{u} &=& |a| \hat{x}_+ + \frac{1}{a} \hat{x}_-,\\
  \hat{v} &=& |a| \hat{p}_+ - \frac{1}{a} \hat{p}_-,
\end{eqnarray}
where $a$ is a non-vanishing real number.  With the definition $\langle \hat{\mathcal O}\rangle = Tr \left[ \hat{\rho}^{(2)} \hat{\mathcal O} \right]$, one computes the fluctuations $\langle(\Delta u)^2\rangle = \langle (\hat{u}-\langle\hat{u}\rangle)^2\rangle$ and $\langle(\Delta v)^2\rangle = \langle (\hat{v}-\langle\hat{v}\rangle)^2\rangle$.  If
\begin{equation}
f(a) \equiv \left( \langle(\Delta u)^2\rangle + \langle(\Delta v)^2\rangle \right)/ \left( a^2 +1/a^2 \right) > 1
\label{criterion}
\end{equation}
for {\it any} choice of $a$, the two modes are inseparable.

\begin{figure}
\includegraphics[width=0.8\textwidth,trim = 0 400 -100 0,clip]{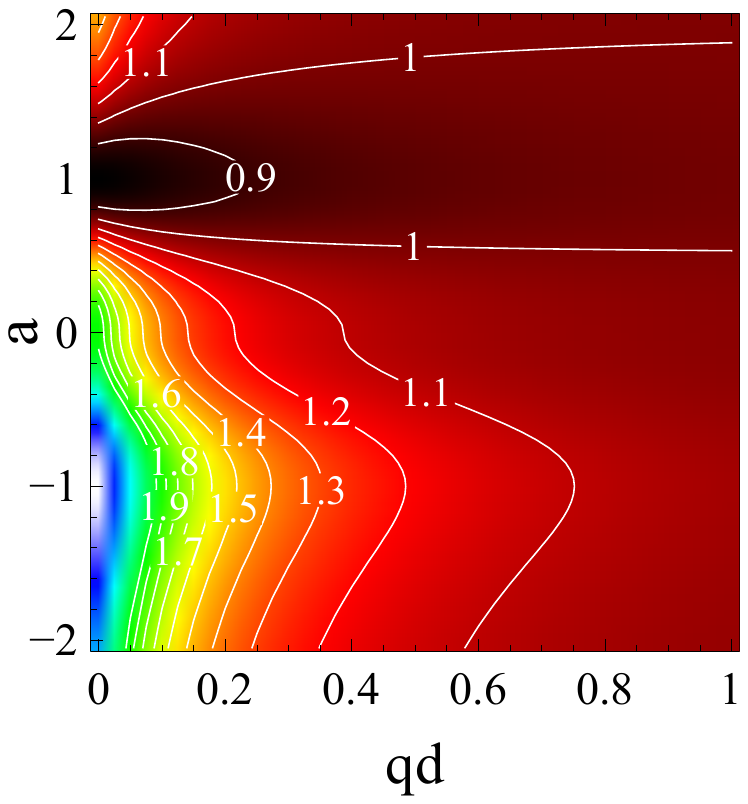}
\caption{Fluctuation factor $f(a)$ as function of $a$ and $qd$ (see text.)  Contour lines indicate level sets of $f(a)$.  If $f(a)>1$ for any $a$, $\pm \qn$ modes at the corresponding $qd$ are inseparable. }
\label{Figure3}
\end{figure}

Figure \ref{Figure3} illustrates the behavior of the fluctuation factor $f(a)$ for different choices of $a$ as a function of $qd$.  One can see clear regions where $f(a) > 1$ for $qd < 1$, particularly for $a \sim -1$ and $a \sim 2$.  This demonstrates that plasmon modes $\pm \qn$ with $q \lesssim 1/d$ will necessarily have entanglement properties.  It should be noted that the criterion in Eq. \ref{criterion} is a sufficient condition for inseparability, but not a necessary one \cite{Duan_2000}.  Thus this equation yields a minimal bound on inseparable modes, but modes outside this region may also by inseparable.  Nevertheless, the analysis shows unequivocally that some of the plasmon modes generated by the procedure leading to $\rho^{(2)}$ will be inseparable.

\begin{figure}
\includegraphics[width=8.cm]{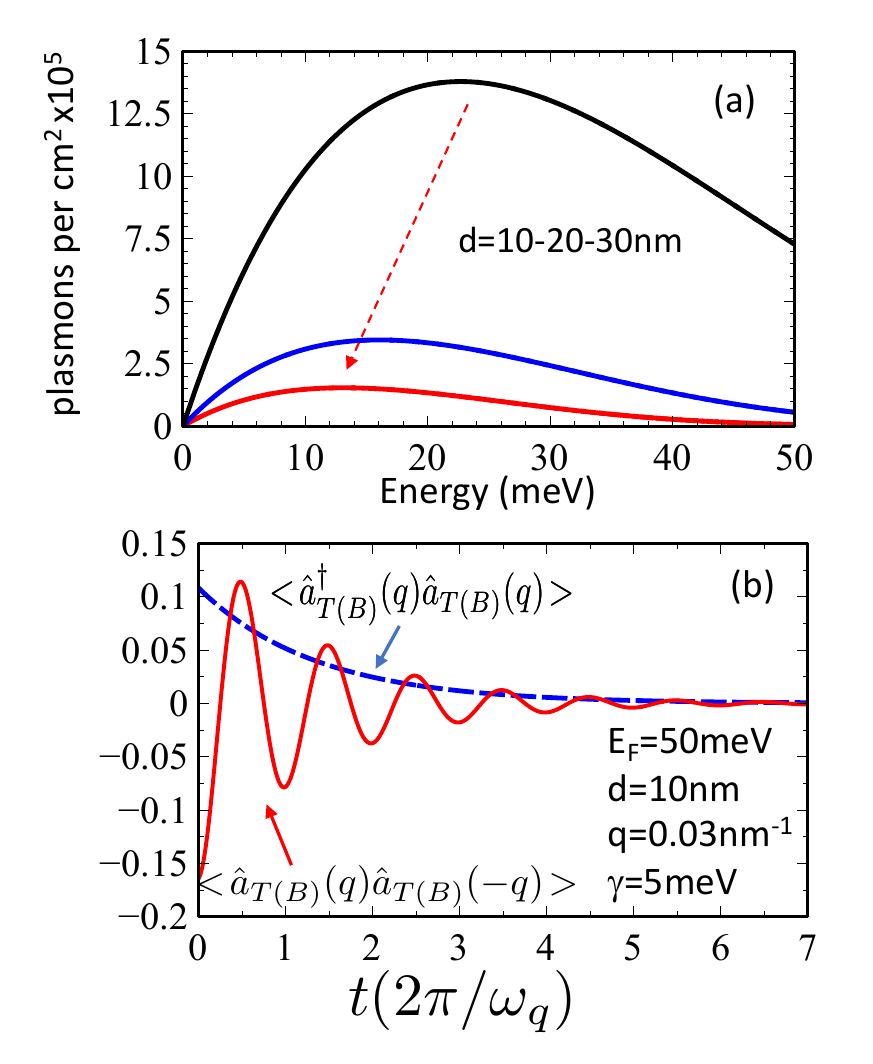}
\caption{(a) (b) Number of plasmon per square centimeter as function of the energy for different values of layer separation  $d$.
Time dependence of the number of plasmons and intralayer coherence in the top layer, after the bottom layer is suddenly disconnected.
Inset in (b) shows the parameters used in the calculation.}
\label{Figure2}
\end{figure}

The density of emitted plasmons at a given energy immediately after the bottom layer depletion depends on the doping of the DL system and on the layer separation. Fig. \ref{Figure2}(a) illustrates this for different values of $d$.  For a Fermi energy of 50meV, dielectric constant $\epsilon_d$=5,
and  $d$=10nm,  the initial population has a maximum  near 10meV and the density of plasmons is of order 10$^{5}$ cm$^{-2}$. This non-monotonic behavior is present because the density of plasmon states vanishes at zero energy, whereas the population of a given mode, $n_{\qn} \equiv \langle a_T^{\dag}({\bf q}) a_T({\bf q})\rangle$, vanishes at high energy.  Because the effective Rabi coupling depends on $qd$, as the layer separation increases the maximum moves to smaller wavevectors and so lower energies.
The
coupling between layers decreases exponentially with $d$ so that the
density of  virtual  plasmons also decreases with increasing $d$.
Interestingly, the density  of virtual plasmons in each layer integrated over all momentum  depends {\it only} on $d$, and is independent of the doping of the layers:  $\frac 1 S \sum _{\qn} n_{\qn} \approx \frac {0.0036} {d^2}$.

After the bottom layer depletion at time $t$=0,
the  time evolution of the density matrix can be modeled using a Markovian master equation,
where dissipation in the top layer is introduced  perturbatively by means of a
Lindblad operator \cite{Leonhardt-book},
\begin{equation}
\partial _t  \hat \rho _{\qn}   =-\frac i {\hbar} [\hat  H_T , \hat \rho _{\qn} ] +
\frac {\gamma} 2   {\cal L} _{\hat a ^T _{\qn}} \, \, ,
\label{equ_mot}
\end{equation}
where the Lindblad operator associated with a bosonic operator $c$ has the form
$ {\cal L}_{c} \equiv 2 c  \hat \rho _{\qn}  c^{ \dagger}  -
c^{ \dagger }    c  \hat \rho _{\qn}
- \hat \rho _{\qn} c ^{\dagger}  c$,
here  $\hat H _T$ is the Hamiltonian of the isolated top layer
and $\gamma$ is the (phenomenological) plasmon decay lifetime that in typical graphene samples is in the range of {\rm meV}\cite{Chen12,Fei12,Woessner:2015aa}.

The initial ($t=0$) form for $\rho$ needed to solve Eq. \ref{equ_mot} is the density matrix of the coupled DL plasmon system in the state $|G\rangle$, traced over plasmon states in the bottom layer.
Fig. \ref{Figure2}(b) plots the number of plasmons in the top layer as a function of time as well as the coherence between oppositely propagating plasmons,
as obtained from the solution to Eq. \ref{equ_mot}. The number of plasmons decays exponentially with rate $\gamma$, as does the coherence, the latter oscillating
with the top layer plasmon frequency $\omega_q$.

\section{
Summary and Discussion}
\label{sec:summary}
\vspace{-0.25cm}
In this paper we have developed the theory of plasmons in a double layer by modeling them as confined electromagnetic modes of the structure, an approach that is particularly appropriate when the electromagnetic coupling between layers is strong.  The formalism reproduces the expected dispersion of the plasmon modes, and moreover allows an exploration of their quantum properties.  An interesting perspective the approach reveals is the presence of virtual plasmons associated with each layer due to quantum fluctuations.  These plasmons are generally entangled, but cannot be released from the ground state without some parametric change in the Hamiltonian.  We consider doing this with a form of the dynamical Casimir effect \cite{Nation_2012}, a sudden depletion of one layer, and find that pairs of plasmons with equal and opposite momenta indeed escape.  Immediately after the depletion, plasmons in the $\pm \qn$ modes are inseparable for $qd \lesssim 1$, and so will support purely quantum correlations in their populations.  Moreover, the modes maintain coherence over a period of time determined by environmental dissipation.

The protocol described above is, for currently available materials, challenging to carry out \cite{Main1}. 
Is there a more practical way to release inseparable plasmons from this system?  One possibility would be to modulate the electron density of one of the two layers at some frequency, and search for plasmons released at half that frequency \cite{Sun:2022aa}; this represents yet another realization of the dynamical Casimir effect.  Beyond this, it is interesting to consider how such released plasmons could be detected
along with their spectral distribution.  One possibility is to search for far infrared narrow-band emission
which results from their radiative decay. This technique has been used  in early work on traditional semiconductor heterostructures \cite{Tsui:1980aa,Hopfel:1982aa,Okisu:1986aa} and more recently in graphene \cite{Li:2019aa}.  Moreover, the inseparability of $\pm \qn$ plasmon modes this system can release in a dynamical Casimir effect protocol should be detectable via correlations in fluctuations of the electric field on either side of the system, which in principle could be detected via near-field microscopy techniques. \cite{Fei12,Chen12,Fei:2013aa} Studies of these possibilities will be addressed in future work.

%\vspace{0.1cm}
%\par \noindent
%{\it Plasmon band structure in  a  patterned layer. }
%!TEX encoding = UTF-8 UnicodIn graphene a periodic array of quantum dots can be created by patterning or gating.
%We describe arrays of dots in graphene as a periodic modulation of the carrier density\cite{Nikitin:2012aa,Nikitin:2014aa}.
% ribbons and graphene dots as isolated regions with a finite Drude weight, thus arrays of dots or ribbons  are described by a periodic modulation of the Drude weight.
%In presence of a modulation of the density, the Hamiltonian for   quantum plasmons in an uniform layer, Eq.\ref{QH},  changes  and the  second term, that corresponds to  the kinetic energy should include the
%Drude weight spatial modulation.\cite{Supp1}
%The periodicity of the perturbation,
%couples plasmons with different momentum and creates a plasmon band structure that is defined
%in a Brillouin zone dictated by the periodicity of the modulation.\cite{Supp1}
%!TEX encoding = UTF-8 UnicodeThe plasmon

\vspace{0.2cm}
\noindent
{\it Acknowledgments}. The authors thank L. Mart\'{\i}n-Moreno, C. Tejedor and R.E. Ferreira for helpful discussions. LB was supported by Grant PID2021-125343NB-I00 (MCIN/AEI/FEDER, EU).  HAF
acknowledges the support of the NSF through Grant
Nos. ECCS-1936406 and DMR-1914451 and of the Vice Provost for Research of Indiana University through the Faculty Research Support Program.

\newpage

\begin{widetext}
\begin{center}
\large{SUPLEMENTARY MATERIAL} \\ \vspace{0.2cm}
\large {Quantum Plasmons in Double Layer Systems  }

\end{center}

\par \noindent
\section{Plasmons as confined light in a single lossless metallic layer.}
\par \noindent
We consider a metallic 2D-layer, placed at the $z=0$ plane and characterized by a pure  imaginary optical conductivity
\begin{equation}
\sigma (\omega) =i \frac D {\omega},
\label{Drude}
\end{equation}
where $D$ is  the Drude weight of the 2D metal.
We look for plasmons as light confined in the 2D-layer.
Above and below the plane the wave equation for the magnetic field of the light is
\begin{equation}
\nabla ^2 {\bf H} - \frac {\varepsilon} {c ^2} \frac {\partial ^2}{\partial t ^2} {\bf H} =0.
\end{equation}
(Note that for $z \neq 0$ there are neither free charges nor currents in the space).
We consider TM waves with the magnetic field perpendicular both to the plane and to the propagation  direction, $ \pm {\hat x}$.
%Now we have to impose boundary condition to the magnetic field.
We look for light confined to the plane, so that the solutions should have the form ($i=1$ referring to $z > 0$, $i=2$ to $z<0$),
\begin{eqnarray}
{\bf H}_1&=&  H_{1,y} e ^{i k_x x} e ^{-\kappa z} e^{-i \omega t}{\hat \yn},  \nonumber \\
{\bf H}_2&=&  H_{2,y} e ^{i k_x x} e ^{\kappa z} e^{-i \omega t}{\hat \yn  },
\end{eqnarray}
with $k_x ^2- \kappa ^2 = \frac {\epsilon} {c^2} \omega^2 $.
In graphene and for the usual doping levels  the speed of light can be considered infinity and we  assume $\kappa$=$|k_x|$.

From Maxwell's equation
\begin{equation}
{\bf \nabla} \times  {\bf H}_i = \frac {\partial {\bf D} _i}{\partial t}
\end{equation}
one finds
\begin{eqnarray}
E_{i,x} & =&- (-1) ^i   \frac {\kappa  H_{i,y}} {\omega \epsilon \epsilon _0} i,  \nonumber \\
E_{i,z} & =&  -\frac { k_x  H_{i,y} } {\omega \epsilon \epsilon _0}.
\end{eqnarray}
We now impose boundary conditions on the electric and magnetic fields,
\begin{eqnarray}
\hat z \times ({\bf E } _2 - {\bf E}_2) &=&0 \, \, \rightarrow  \, E_{1,x} (z=0)=E_{2,x}(z=0)=E_x(z=0) \, \, \rightarrow H_{1,y}=-H_{2,y},
 \nonumber  \\
\hat z  \times ({\bf H} _2- {\bf H} _1)&  = &-  {\bf J} =-\sigma {\bf E} \, \rightarrow \, H_{2,y}(z=0)-H_{1,y}(z=0)=-J_x = -\sigma E_x(z=0),
\end{eqnarray}
to obtain the self consistent equation for the existence of  plasmons,
\begin{equation}
2 \! =\! -i \frac {\sigma \kappa}{\omega \epsilon \epsilon_0}. \,
\end{equation}
In the case of a Drude-like lossless system the plasmon dispersion becomes
\begin{equation}
\omega  \! =\!\sqrt { \frac { D}{2 \epsilon _0 \epsilon} \kappa}.  \,
\end{equation}
and for graphene the Drude weight is related to the Fermi energy through
\begin{equation}
D \! =\! \frac {e ^2 E_F}{\hbar ^2  \pi} \, .
\end{equation}
Moreover, for  graphene the Fermi energy is related with the density through the relation
\begin{equation}
E_F=\hbar v_F\sqrt{\pi n_0 }\, ,
\end{equation}
where $v_F$ is the graphene Dirac velocity and  $n_0$ is the density of carriers in the layer.

Finally, for a general in-plane propagation direction the electric field associated with the plasmon takes the form
\begin{equation}
{\bf E}_{\qn}(\rn,t)= E_0 e ^{i \qn  \rn } e ^{-\kappa |z|}  e^{-i\omega t } \left (i \frac {\qn } q -  \frac {z}{|z|}
 \hat {\zn} \right ).
\end{equation}
In graphene and for typical doping levels, the plasmon wavelength is much larger than the wavelength of light at the same frequency, for frequencies where plasmons are usually detectable.  This justifies our use of the
semi-static limit, in which $\kappa$=$q$.

\section{Quantization of Plasmons in a  Drude-like lossless graphene layer}
The real-valued vector potential and electric fields of associated with a plasmon mode have the form
\begin{eqnarray}
{\bf A}( \rn,t) & = &  \sum _{\qn} A_{\qn}   e^{i \qn \rn} e^{-i \omega _q t } {\bf u}(\qn,\kappa,z) + c.c.,
\nonumber \\ {\bf E}( \rn,t) & = & \sum _{\qn}  E_{\qn}   e^{i \qn \rn} e^{-i \omega_q t } {\bf u}(\qn,\kappa,z) + c.c.,
\label{vecpot}
\end{eqnarray}
with  $A_{\qn} =-\frac {i}{\omega _q} E_{\qn} $
%and  $ \alpha _{\qn}$ is a constant to be determined \textbf{($\alpha_q$ does not appear in any of the expressions)} 
and
\begin{equation}
{\bf u}(\qn,\kappa,z )= \frac {e^{-\kappa |z|}} {\sqrt {L(q,\kappa)} }\left (i \frac {\qn } q -  \frac {z}{|z|}
 \hat {\zn} \right ),
\end{equation}
with the quantity $L(q,\kappa)$ chosen to normalize the $z$-component of the vector potential,
\begin{equation}
L(q,\kappa)= \frac 1 {\kappa} \left ( 1+\frac {q ^2}{\kappa ^2}\right ).
\end{equation}
The time-averaged energy of the electromagnetic field associated with the plasmons is then
%(Landau-Lifshitz, Brillouin)
\begin{eqnarray}
U & = &    \frac {  \epsilon _0} 2      \int  \int d \rn dz |\En (\rn,t=0)|^2   \frac { d(\omega \epsilon (\omega))}{d \omega}+
\frac 1 2 \int d \rn dz \frac 1 {\mu _0} |{\bf B}(\rn,t=0)|^2 \nonumber \\
& = &        { \epsilon _0}  S \sum _{\qn}   \int dz |E_{\qn}|^2   \frac { d(\omega \epsilon (\omega))}{d \omega} |{\bf u}(\qn,\kappa,z )|^2 +
\frac 1 2 \int d \rn dz \frac 1 {\mu _0} |{\bf B}(\rn,t=0)|^2,
\label{Ufull}
\end{eqnarray}
where $S$ is the surface area of the graphene two-dimensional sheet, and
\begin{equation}
{\bf B}(\rn,t )=  {\bf {\nabla} }\times {\bf A} (\rn,t).
\end{equation}
We model the graphene sheet as a pure 2D material characterized by an optical conductivity.  In this situation the dielectric constant of the system, $\epsilon (\omega)$,  has the form 
\begin{equation}
\epsilon(\omega)=\epsilon _d +\frac i {\epsilon _0 \omega}\sigma (\omega) \delta (z), \, \,
\end{equation}
where $\epsilon _d$ is the dielectric constant of the medium surrounding the graphene layer and
 $\sigma (\omega)$ is the graphene  optical conductivity.  
At low frequencies we assume the latter to be local and lossless, taking the form
\begin{equation}
\sigma (\qn,\omega)=i \frac {D}{\omega}.
\end{equation}
In this situation,
\begin{equation}
\frac { d(\omega \epsilon (\omega))}{d \omega}=\epsilon _d  +\frac D{\omega ^2} \frac 1{\epsilon _0} \delta (z).
\label{dielectric}
\end{equation}
\subsection{Magnetic contribution to the energy.}
The magnetic field is computed from the vector potential using
\begin{equation}
{\bf {\nabla} }\times {\bf A} (\rn,t=0)= \sum _{\qn} i A_{\qn}  e^{i \qn \rn} \frac {e^{-\kappa |z|}}{\sqrt {L(k,\kappa)}} \frac {q^2 -\kappa^2}{\kappa} \frac z {|z|} \left (
\hat \zn \times \frac {\qn} q \right )+ c.c.,
\end{equation}
so that
\begin{equation}
U_M= \frac 1 {2 \mu  _0 }   S \sum _{\qn}
\left (   \frac { q^2 - \kappa ^2  } {\kappa} \right ) ^2
\frac {\kappa ^2}{\kappa ^2 + q^2}
\left  (
A_{\qn}    {A_{\qn} } ^ *   +
{A_{\qn} } ^ *   A_{\qn}
\right ).
\end{equation}
As mentioned in the main text, we adopt the semi-static limit, for which $q \rightarrow \kappa$.  Under this assumption,
\begin{equation}
U_M=0.
\end{equation}

\subsection{Electric contribution}
Using the above result, as well as Eqs. \ref{Ufull} and \ref{dielectric}, we obtain an effective energy functional
%\begin{equation}
%U =      \frac { \epsilon _0} 2  S \sum _{\qn} \omega_{\qn}  ^2   \int dz |A_{\qn}|^2
%\frac 1 2 S \left ( \epsilon _d +\frac D {\omega_q  ^2 } \delta (z) \right )   |{\bf u}(\qn,\kappa,z )|^2
%\label{uem}
%\end{equation}

\begin{equation}
U =       { \epsilon _0 \epsilon _d }  S \sum _{\qn}  | E _{\qn}|^2    \int dz    |{\bf u}(\qn,\kappa,z )|^2 +
  S D \sum _{\qn}    | A _{\qn}|^2  \int dz
  \delta (z)     |{\bf u}(\qn,\kappa,z )|^2
\label{uem}
\end{equation}
In this form we interpret the first term as the energy of the confined light and the second as the kinetic energy of the carriers in the
graphene layer.
Performing the real space integrals, and using the relations anong $D$, $q$ and $\omega _q$, as well as the relation between $E _{\qn}$ and $A _{\qn}$, we arrive
at the simple relation
\begin{equation}
U=2 \epsilon _0 S \epsilon _d \sum_{\qn}  A_{\qn}^*A_{\qn}.
\end{equation}

To quantize this,  we use the correspondence of the classical energy
with that of a set of harmonic oscillators.  The quantized Hamiltonian can be written in the form
\begin{equation}
\hat H=\sum_{\qn} \frac {\hbar \omega _q } 2 \left [ \hat a _{\qn} \hat a ^{\dagger}_{\qn} + \hat a ^{\dagger}_{\qn} \hat a _{\qn} \right ]
\label{HSHO}
\end{equation}
where we have mapped the coefficients
\begin{eqnarray}
A_{\qn}   &\rightarrow  & -i \sqrt{ \frac {\hbar}{ 2 \epsilon _0 \epsilon_d S \omega_q}} \hat a _{\qn}, \nonumber \\
{A_{\qn} }  ^* & \rightarrow & i \sqrt{ \frac {\hbar}{ 2 \epsilon _0  \epsilon_d  S \omega_q }} \hat a ^{\dagger} _{\qn}.
\end{eqnarray}
The bosonic operator $\hat a ^{\dagger}_{\qn}$ creates a plasmon with momentum $\qn$.

In terms of $\hat a _{\qn} $ and $ \hat a ^{\dagger}_{\qn} $, the electric field and vector potential operators become
\begin{eqnarray}
{\hat \En} (\rn,z)  &
=&   \sqrt{ \frac {\hbar \omega _q }{ 2 \epsilon _0 \epsilon_d S  }}   e ^{i \qn  \rn } \un (\qn,z)  \hat a _{\qn} + h.c., \nonumber \\
{\hat \An} (\rn,z) & = &  -i \sqrt{ \frac {\hbar }{ 2 \epsilon _0 \epsilon_d S \omega _q  }}   e ^{i \qn  \rn } \un (\qn,z)  \hat a _{\qn} + h.c.,
\end{eqnarray}
and the quantum Hamiltonian, Eq. \ref{HSHO}, is equivalent to (again, in the quasi-static limit)
\begin{equation}
\hat H   =     \frac {  \epsilon _0 \epsilon _d} 2      \int  \int d \rn dz   {\hat \En} (\rn,z) {\hat \En} (\rn,z)  +
\frac 1 2   \int  \int d \rn dz   D \delta (z)    {\hat \An} (\rn,z) {\hat \An} (\rn,z).
\label{QH}
\end{equation}

\section{Plasmons in a Double Layer: Classical result}

We consider  two parallel 2D metallic layers located at $z=d/2$ (top layer) and at $z=-d/2$ (bottom layer).  Both layers  are loss-less and are characterized   by optical conductivities $\sigma ^{T(B)} (\omega)= i\frac {D^{T(B)}}{\omega} $.
The current in each layer is proportional to the sum of the external field $\En_{ext}$ and the fields created by charge density modulations induced
by the currents in the top and bottom layers, $\En_{ind}^{T}$ and $\En_{ind}^{B}$, respectively. This leads to self-consistent equations for the currents,
\begin{eqnarray}
\Jn ^{T(B)}(\qn) & = & \sigma ^{T(B)} \left [\En _{Ext}(\qn) +\En ^{T(B)} _{ind} (\qn)\right ]= \sigma  ^{T(B)} \left [\En _{Ext}(\qn) -i \qn \phi^{T(B)}  _{ind} (\qn)\right ] \nonumber \\
 & = & \sigma ^{T(B)} \left [\En _{Ext}(\qn) -i \qn v _q \left ( \delta \rho  (\qn)^{T(B)} +e ^{-q d} \delta \rho (\qn)^{B(T)} \right ) \right ] \nonumber \\
 & =& \sigma ^{T(B)} \left [\En _{Ext}(\qn)-i \qn v _q \frac {\qn}{\omega} \left ( \Jn ^{T(B)} +e^{-qd} \Jn ^{B(T)} \right ) \right ]
  \end{eqnarray}
In these equations, $\phi^{T(B)}_{ind}$ is the potential associated with the induced electric fields in the top (bottom) layer, $v_q$=$\frac {e ^2}{2 \epsilon_0 \epsilon_d q}$ is the two dimensional Fourier transform of the Couloumb interaction between charge modulations in the same layer, and $v_q e^{-qd}$ is the Coulomb interaction between
charge modulations in opposite layers.  These modulations are expressed as 
$\delta\rho ^{T(B)} (\qn)$ for the top (bottom) layer. Current densities and charge densities  are related by the continuity equation,
$\qn \Jn ^{T(B)}$=$\omega  \delta \rho  (\qn)^{T(B)}$.

In the absence of external fields the currents satisfy
\begin{equation}
\left ( \begin{array}{c}
\Jn ^{T} \\
\Jn ^B
\end{array} \right ) =
\left ( \begin{array}{cc}
\frac {q^2}{\omega ^2 } v(q) D ^T &\frac {q^2}{\omega ^2 } v(q) D ^B  \\
\frac {q^2}{\omega ^2 } v(q) D ^B &\frac {q^2}{\omega ^2 } v(q) D ^T
\end{array}
\right )
\left ( \begin{array}{c}
\Jn ^{T} \\
\Jn ^B
\end{array} \right ) =
\left ( \begin{array}{cc}
\frac {\omega _T ^2  (q) }{\omega ^2} & \frac  {\omega   _B ^2 (q)  } {\omega ^2}  e ^{-qd}\\
 \frac {\omega _T ^2 (q) }{\omega ^2}e ^{-qd} &  \frac {\omega _B ^2 (q)  }{\omega ^2}\end{array}
\right )
\left ( \begin{array}{c}
\Jn ^{T} \\
\Jn ^B
\end{array} \right ),
\end{equation}
where $\omega _{T(B)} (q) $ is the plasmon frequency of the top (bottom) layer in isolation.
Self-sustained plasmons occur at the zeros of the determinant of the matrix
\begin{equation}
\left ( \begin{array}{cc}
1-\frac {\omega _T ^2 (q) }{\omega ^2} & -\frac  {\omega _B ^2 (q) } {\omega ^2} e^{-qd} \\
-\frac {\omega _T ^2 (q) }{\omega ^2}e^{-qd}& 1- \frac {\omega _B ^2(q)  }{\omega ^2}\end{array}
\right ),
\end{equation}
which occur when
\begin{equation}
\omega ^2 = \omega_{\pm}^2(q) \equiv \frac {(\omega _T ^2 (q) +\omega _B ^2 (q) ) \pm \sqrt {(\omega _T ^2( q) +\omega_B^2 (q) ) ^2 -4 \omega _T^2 (q) \omega _B^2 (q) (1-e^{-2 q d})}} 2.
\label{DLPlasmons}
\end{equation}
In the limit of small momentum the two plasmon modes become optical and acoustic, with forms
\begin{eqnarray}
\omega_+(q) &  \stackrel{ qd \rightarrow 0}{\rightarrow}  & \sqrt {\omega_T ^2 (q)  +\omega _B ^2 (q) }, \nonumber \\
\omega_-(q) &  \stackrel{ qd \rightarrow 0}{\rightarrow}  & \sqrt { 2 \frac {\omega _T ^2(q) \omega _B ^2 (q)} {\omega_T ^2 (q)+\omega _B ^2 (q) } q d}.
\nonumber \\\label{omega_classical}
\end{eqnarray}
Both $\omega _T$ and $\omega _B$ disperse as $\sqrt  q$, so that at long wavelength  the optical plasmon $\omega_+(q) \sim \sqrt{q}$, and the acoustic mode $\omega_-(q) \sim q$.
%In order to obtain these results we solve self-consistently the electric fields and currents on both layers.

\section{Quantum Coupling Between Plasmons in Parallel Layers}
The electric  field operators associated with the top layer located at $z$=$d/2$ and the bottom layer located at $z$=$-d/2$ are
\begin{eqnarray}
\hat {{\En}}^T ( \rn) & =&   \sum _{\qn} \sqrt{ \frac {\hbar}{ 2 \epsilon _0 \epsilon S \omega _T (q) }} e^{i \qn \rn}  {\bf u}(\qn,\kappa,z \! - \! \frac d 2  )  \, \hat a ^T _{\qn} +h.c.,  \nonumber \\
\hat {{\En}}^B( \rn) & =& \sum _{\qn} \sqrt{ \frac {\hbar}{ 2 \epsilon _0 \epsilon S \omega _B (q) }} e^{i \qn \rn}  {\bf u}(\qn,\kappa,z \! +\! \frac d 2  )  \, \hat a ^B _{\qn} +h.c.
\end{eqnarray}
The total electric field operator is
\begin{equation}
{\hat \En}(\rn)=\hat {{\bf E}}^T ( \rn)+\hat {{\bf E}}^B ( \rn).
\end{equation}
Following steps analogous to those leading to Eqs. \ref{HSHO} and \ref{QH}, the quantum Hamiltonian of the double layer system becomes
\begin{equation}
\hat H= \hat H_T+\hat H_B+\Delta {\hat H},
\label{HDL}
\end{equation}
with
\begin{eqnarray}
\hat H_T & =&  \sum _{\qn }  \frac {\hbar \omega _T }2 \left ( \hat a ^T   _{\qn} { } ^{\dagger}  \, \hat a ^T   _{\qn} +\hat a ^T _{  \qn } \, \hat a ^T  _{\qn} {} ^{\dagger}   \right ),\nonumber \\
\hat H_B & =&  \sum _{\qn }  \frac {\hbar \omega _B }2 \left ( \hat a ^B  _{ \qn} { } ^{\dagger} \, \hat a ^B _  {\qn} +\hat a ^B _{ \qn }\,  \hat a ^B _{\qn}{ } ^{\dagger}   \right ).
\end{eqnarray}
%
%
%
%
%
%
%
%
%\begin{comment}
The coupling between plasmon degrees of freedom in the top and bottom layers appears in the last term, and arises due to the cross term in squaring the electric field to form the energy.  Its concrete form is
\begin{equation}
\Delta {\hat H}  = \epsilon _0 \epsilon \int dV \,  {\hat {{\bf E}}^T ( \rn,z)} {\hat {{\bf E}}^B ( \rn,z) }=
\sum _{\qn} \hbar \frac { \sqrt{\omega _T \omega _B}} 2 e^{-q d}
\left (   \hat a ^T   _{\qn} { } ^{\dagger}  \, \hat a ^B   _{\qn}  + \hat a ^B   _{\qn} { } ^{\dagger}  \, \hat a ^T   _{\qn}
- \hat a ^T   _{\qn}   \, \hat a ^B   _{-\qn}  - \hat a ^B   _{\qn} { } ^{\dagger}  \, \hat a ^T   _{-\qn} {}^{\dagger}
\right ),
\label{DLC}
\end{equation}
%& = & \sum _{\qn} \hbar \frac { \sqrt{\omega _T \omega _B}} 4 e^{-q d}
%\left ( \hat a   _{\qn}  ^{T \dagger}  \, \hat a ^B   _{\qn} -\hat a   _{\qn}  ^{T \dagger}  \,\hat a    _{-\qn} ^{B \dagger}
%+\hat a   _{\qn}  ^{B \dagger}  \, \hat a ^T   _{\qn} -\hat a   _{\qn}  ^{B \dagger}  \,\hat a    _{-\qn}  ^{T \dagger}
%-\hat a ^T   _{-\qn} { }  \hat a ^B   _{\qn} +\hat a ^T   _{-\qn} { }  \hat a    _{-\qn} ^{B \dagger}
%-\hat a ^B  _{-\qn} { }  \hat a ^T   _{\qn} +\hat a ^B   _{-\qn} { }  \hat a    _{-\qn} ^{T \dagger} \right )
%\end{eqnarray}
in which we see that, for every $\qn$, modes at $\pm\qn$ from both the top and bottom layer are admixed.

\section{Diagonalization of the double layer (DL) plasmon Hamiltonian}
In matrix form the Hamiltonian Eq. \ref{HDL} can be written as
\begin{equation}
{\hat H}= \frac {\hbar} 2 \sum _{\qn  }
\left (
\hat a _ T ^{\dagger} (\qn ),
\hat a _ B  ^{\dagger}(\qn ),
\hat a _ T  (-\qn ),
\hat a _ B  (-\qn )
\right )
\left ( \begin{array}{cccc}
\omega _T& \wqd & 0  &  -\wqd  \\
\wqd & \omega _B   &-\wqd &0   \\
0 &- \wqd   &\omega _T &\wqd  \\
-\wqd & 0 & \wqd & \omega _B
\end{array} \right )
\left ( \begin{array} {c}
\hat a _ T  (\qn ) \\
\hat a _ B  (\qn )\\
\hat a _ T  ^{\dagger} (-\qn ) \\
\hat a _ B  ^{\dagger} (-\qn)
 \end{array}
\right ).
\end{equation}
This Hamiltonian is bilinear in the field operators and can be diagonalized through a Bogoliubov   transformation.  Following the pioneering work of Hopfield, we introduce the operator transformation
\begin{equation}
\left ( \!  \! \! \begin{array} {c}
 \hat b_1  (\qn) \\
\hat  b_2  (\qn)\\
\hat  b_1   ^{\dagger} (-\qn) \\
\hat  b_2  ^{\dagger} (-\qn)
 \end{array} \!
\right )=U ^{-1} \left ( \!  \! \! \begin{array} {c}
 \hat a _ T  (\qn) \\
\hat a _ B (\qn)\\
\hat a _ T  ^{\dagger} (-\qn) \\
\hat a _ B ^{\dagger} (-\qn)
 \end{array} \!
\right )
=\left ( \begin{array}{cccc}
w_{1,\qn}  & x_{1,\qn} & y _{1,\qn} & z_{1,\qn }  \\
w_{2,\qn}  & x_{2,\qn} & y _{2,\qn} & z_{2,\qn }  \\
w_{3,\qn}    & x_{3,\qn}   & y _{3,\qn}   & z_{3,\qn }    \\
w_{4,\qn}    & x_{4,\qn}   & y _{4,\qn}   & z_{4,\qn }
\end{array} \right ) \! \! \!
\left ( \!  \! \! \begin{array} {c}
 \hat a _ T  (\qn) \\
\hat a _ B (\qn)\\
\hat a _ T  ^{\dagger} (-\qn) \\
\hat a _ B  ^{\dagger} (-\qn)
 \end{array} \!
\right ).
\end{equation}
For the various entries in the $4\times 4$ matrix,
index 1 corresponds to the annihilation operator of the lower positive energy mode, index 2 to the annihilation operator of the larger positive energy mode, and indices 3 and 4 correspond to creation operators for the 1 and 2 modes, respectively. %$E_3$=-$E_1$ and $E_4$=$-E_2$.
We require the transformation to preserve the bosonic commutations relations obeyed by the $a_{T,B}(\qn)$, $a_{T,B}^{\dag}(\qn)$ operators, which leads to the conditions
\begin{eqnarray}
w_{i,\qn}w^*_{i',\qn}+x_{i,\qn}x^*_{i',\qn}-y_{i,\qn}y^*_{i',\qn}-z_{i,\qn}z^*_{i',\qn} &=&\delta_{i,i'}  \, \, \, {\rm for} \, \, i=1,2 \, , \nonumber \\
w_{i,\qn}w^*_{i',\qn}+x_{i,\qn}x^*_{i',\qn}-y_{i,\qn}y^*_{i',\qn}-z_{i,\qn}z^*_{i',\qn} &=& -\delta_{i,i'} \, \, \, {\rm for}\, \,  i=3,4.
\label{normalization}
\end{eqnarray}
These relations imply
\begin{equation}
U=
\left ( \begin{array}{cccc}
w_{1,\qn} ^*   & w_{2,\qn}^*  & -w _{3,\qn} ^* & -w_{4,\qn } ^* \\
x_{1,\qn}  ^* & x_{2,\qn} ^* & -x _{3,\qn} ^* & -x_{4,\qn } ^* \\
-y_{1,\qn} ^*   & -y_{2,\qn} ^*   & y _{3,\qn} ^*   & y_{4,\qn } ^*    \\
-z_{1,\qn}  ^*  & -z_{2,\qn} ^*   & z _{3,\qn}  ^*  & z_{4,\qn } ^*
\end{array} \right ).
\end{equation}

The matrix $U^{-1}$ is related with  the  adjoint of $U$ through the relation
\begin{equation}
U^{\dagger} = \Gamma U ^{-1} \Gamma,
\, \, \,\,  {\rm with}\, \, \, \, \Gamma=
\left ( \begin{array}{ccrr}
1  & 0  & 0 &0 \\
0  & 1  & 0 & 0 \\
0  & 0  & -1 & 0    \\
0  & 0  & 0  &-1
\end{array} \right ).
\end{equation}
The transformed Hamiltonian can then be written
\begin{equation}
\hat H =
\left ( \hat b_1  ^{\dagger} (\qn) ,
\hat  b_2  ^{\dagger} (\qn),
\hat  b_1    (-\qn),
\hat  b_2  (-\qn) \right ) U^{\dagger} H U \left ( \!  \! \! \begin{array} {c}
 \hat b_1  (\qn) \\
\hat  b_2  (\qn)\\
\hat  b_1   ^{\dagger} (-\qn) \\
\hat  b_2  ^{\dagger} (-\qn)
 \end{array} \!
\right )=
\left ( \hat b_1  ^{\dagger}(\qn) ,
\hat  b_2  ^{\dagger}(\qn),
\hat  b_1   (-\qn),
\hat  b_2   (-\qn) \right ) \Gamma U^{-1} \Gamma H U \left ( \!  \! \! \begin{array} {c}
 \hat b_1  (\qn) \\
\hat  b_2  (\qn)\\
\hat  b_1   ^{\dagger} (-\qn) \\
\hat  b_2  ^{\dagger} (-\qn)
 \end{array} \!
\right ).
\end{equation}
The transformed Hamiltonian is diagonal if and only if the
matrix $U$ diagonalizes  the so-called Hopfield matrix
\begin{equation}
M=\Gamma H= \frac {\hbar \omega _T} 2
\left ( \begin{array}{cccc}
1& \sqd & 0  &  -\sqd  \\
\sqd &\frac{ \omega _B }{\omega _T}  &-\sqd &0   \\
0 & \sqd   &-1 &-\sqd  \\
\sqd & 0 & -\sqd &     -\frac{ \omega _B }{\omega _T} \end{array} \right ),
\end{equation}
which has eigenvalues
\begin{equation}
\omega^{\pm}  = \sqrt {\frac {(\omega _T ^2 +\omega _B ^2) \mp \sqrt {(\omega _T ^2+\omega_B^2) ^2 -4 \omega _T^2 \omega _B^2(1-e^{-2 q d})}} 2}.
%\label{DLPlasmons}
\end{equation}
These coincide with the double layer plasmon frequencies obtained a above in the classical analysis.

In the symmetric case, $\omega_B (q) =\omega_T (q) =\omega (q)  $, and the two normal mode frequencies become
\begin{equation}
\omega^{1(2)}(q)  = \omega (q) \sqrt { 1\mp e ^{-qd}}.
\end{equation}
%the coefficients $w$, $x$, $y$ and $z$ take  the form of U, is
%\begin{equation}
%U=
%\left ( \begin{array}{cccc}
%\frac {-x+1-\sqrt{1-2x}} x & \frac {1-x+\sqrt{1+2x}} x & -1 & 1 \\
%-\frac {-1+x+\sqrt{1-2x}} x & \frac {-1-x-\sqrt{1+2x}} x &1 & 1 \\
%-1& 1 & \frac {x-1-\sqrt{1-2x}} x  & \frac {-1-x-\sqrt{1+2x}} x \\
%1& 1 & -\frac {x-1-\sqrt{1-2x}} x& \frac {-1-x-\sqrt{1+2x}} x
%\end{array} \right )
%\end{equation}
%with $x=\frac 1 2 e ^{-q d}$
%We should normalize these eigenvectors following the relation Eq.\ref{normalization}
The corresponding $U$ matrix takes the form
\begin{equation}
U=\frac 1{\sqrt  2}
\left ( \begin{array}{cccc}
\cosh \theta _1 & \cosh {\theta_2}& \sinh {\theta _1} & \sinh {\theta_2} \\
-\cosh \theta _1 & \cosh {\theta_2}& -\sinh {\theta _1} & \sinh {\theta_2} \\
-\sinh \theta _1 & -\sinh {\theta_2}& -\cosh {\theta _1} & -\cosh {\theta_2} \\
\sinh \theta _1 & -\sinh {\theta_2}& \cosh {\theta _1} & -\cosh {\theta_2}
\end{array} \right ),
\end{equation}
with
\begin{equation}
\tanh {\theta _i}= \frac {\omega (q) -\omega ^i(q)}{ \omega (q)+\omega ^i(q) }
\label{thetas}
\end{equation}
which is equivalent to ${\omega^i(q) }/{ \omega (q)  }=e ^{-2 \theta_i}$.
%\bibliography{ref}
The resultant  Hamiltonian has the diagonal form
\begin{equation}
\hat H = \sum _{\qn, i }
\frac {\hbar {\omega ^i(q)}}2   \left (\hat  b _ i (q)  \hat b_i  ^{\dagger}  (q)  + \hat b _ i ^{\dagger} (q) \hat b _ i (q)  \right ).
\end{equation}
%where the operators $\hat b1$ and $\hat b_2 $ are  operators that annihilate  plasmons with momentum $q$  with frequency  $\omega ^1 $ and $\omega  ^2 $ respectively.

\end{widetext}


%\bibliography{mia}

\begin{thebibliography}{75}%
\makeatletter
\providecommand \@ifxundefined [1]{%
 \@ifx{#1\undefined}
}%
\providecommand \@ifnum [1]{%
 \ifnum #1\expandafter \@firstoftwo
 \else \expandafter \@secondoftwo
 \fi
}%
\providecommand \@ifx [1]{%
 \ifx #1\expandafter \@firstoftwo
 \else \expandafter \@secondoftwo
 \fi
}%
\providecommand \natexlab [1]{#1}%
\providecommand \enquote  [1]{``#1''}%
\providecommand \bibnamefont  [1]{#1}%
\providecommand \bibfnamefont [1]{#1}%
\providecommand \citenamefont [1]{#1}%
\providecommand \href@noop [0]{\@secondoftwo}%
\providecommand \href [0]{\begingroup \@sanitize@url \@href}%
\providecommand \@href[1]{\@@startlink{#1}\@@href}%
\providecommand \@@href[1]{\endgroup#1\@@endlink}%
\providecommand \@sanitize@url [0]{\catcode `\\12\catcode `\$12\catcode
  `\&12\catcode `\#12\catcode `\^12\catcode `\_12\catcode `\%12\relax}%
\providecommand \@@startlink[1]{}%
\providecommand \@@endlink[0]{}%
\providecommand \url  [0]{\begingroup\@sanitize@url \@url }%
\providecommand \@url [1]{\endgroup\@href {#1}{\urlprefix }}%
\providecommand \urlprefix  [0]{URL }%
\providecommand \Eprint [0]{\href }%
\providecommand \doibase [0]{http://dx.doi.org/}%
\providecommand \selectlanguage [0]{\@gobble}%
\providecommand \bibinfo  [0]{\@secondoftwo}%
\providecommand \bibfield  [0]{\@secondoftwo}%
\providecommand \translation [1]{[#1]}%
\providecommand \BibitemOpen [0]{}%
\providecommand \bibitemStop [0]{}%
\providecommand \bibitemNoStop [0]{.\EOS\space}%
\providecommand \EOS [0]{\spacefactor3000\relax}%
\providecommand \BibitemShut  [1]{\csname bibitem#1\endcsname}%
\let\auto@bib@innerbib\@empty
%</preamble>
\bibitem [{\citenamefont {Pines}\ and\ \citenamefont
  {Bohm}(1952)}]{Pines:1952aa}%
  \BibitemOpen
  \bibfield  {author} {\bibinfo {author} {\bibfnamefont {D.}~\bibnamefont
  {Pines}}\ and\ \bibinfo {author} {\bibfnamefont {D.}~\bibnamefont {Bohm}},\
  }\href {\doibase 10.1103/PhysRev.85.338} {\bibfield  {journal} {\bibinfo
  {journal} {Physical Review}\ }\textbf {\bibinfo {volume} {85}},\ \bibinfo
  {pages} {338} (\bibinfo {year} {1952})}\BibitemShut {NoStop}%
\bibitem [{\citenamefont {Pines}(1956)}]{Pines:1956aa}%
  \BibitemOpen
  \bibfield  {author} {\bibinfo {author} {\bibfnamefont {D.}~\bibnamefont
  {Pines}},\ }\href {\doibase 10.1103/RevModPhys.28.184} {\bibfield  {journal}
  {\bibinfo  {journal} {Reviews of Modern Physics}\ }\textbf {\bibinfo {volume}
  {28}},\ \bibinfo {pages} {184} (\bibinfo {year} {1956})}\BibitemShut
  {NoStop}%
\bibitem [{\citenamefont {Sawada}\ \emph {et~al.}(1957)\citenamefont {Sawada},
  \citenamefont {Brueckner}, \citenamefont {Fukuda},\ and\ \citenamefont
  {Brout}}]{Sawada:1957aa}%
  \BibitemOpen
  \bibfield  {author} {\bibinfo {author} {\bibfnamefont {K.}~\bibnamefont
  {Sawada}}, \bibinfo {author} {\bibfnamefont {K.~A.}\ \bibnamefont
  {Brueckner}}, \bibinfo {author} {\bibfnamefont {N.}~\bibnamefont {Fukuda}}, \
  and\ \bibinfo {author} {\bibfnamefont {R.}~\bibnamefont {Brout}},\ }\href
  {\doibase 10.1103/PhysRev.108.507} {\bibfield  {journal} {\bibinfo  {journal}
  {Physical Review}\ }\textbf {\bibinfo {volume} {108}},\ \bibinfo {pages}
  {507} (\bibinfo {year} {1957})}\BibitemShut {NoStop}%
\bibitem [{\citenamefont {Giuliani}\ and\ \citenamefont
  {Vignale}(2005)}]{Vignale-book}%
  \BibitemOpen
  \bibfield  {author} {\bibinfo {author} {\bibfnamefont {G.}~\bibnamefont
  {Giuliani}}\ and\ \bibinfo {author} {\bibfnamefont {G.}~\bibnamefont
  {Vignale}},\ }\href@noop {} {\emph {\bibinfo {title} {Quantum Theory of the
  Electron Liquid}}}\ (\bibinfo  {publisher} {Cambridge University Press},\
  \bibinfo {year} {2005})\BibitemShut {NoStop}%
\bibitem [{\citenamefont {Pitarke}\ \emph {et~al.}(2007)\citenamefont
  {Pitarke}, \citenamefont {Silkin}, \citenamefont {Chulkov},\ and\
  \citenamefont {Echenique}}]{Pitarke:2007aa}%
  \BibitemOpen
  \bibfield  {author} {\bibinfo {author} {\bibfnamefont {J.~M.}\ \bibnamefont
  {Pitarke}}, \bibinfo {author} {\bibfnamefont {V.~M.}\ \bibnamefont {Silkin}},
  \bibinfo {author} {\bibfnamefont {E.~V.}\ \bibnamefont {Chulkov}}, \ and\
  \bibinfo {author} {\bibfnamefont {P.~M.}\ \bibnamefont {Echenique}},\ }\href
  {\doibase 10.1088/0034-4885/70/1/R01} {\bibfield  {journal} {\bibinfo
  {journal} {Reports on Progress in Physics}\ }\textbf {\bibinfo {volume}
  {70}},\ \bibinfo {pages} {1} (\bibinfo {year} {2007})}\BibitemShut {NoStop}%
\bibitem [{\citenamefont {Ritchie}(1957)}]{Ritchie:1957aa}%
  \BibitemOpen
  \bibfield  {author} {\bibinfo {author} {\bibfnamefont {R.~H.}\ \bibnamefont
  {Ritchie}},\ }\href {\doibase 10.1103/PhysRev.106.874} {\bibfield  {journal}
  {\bibinfo  {journal} {Physical Review}\ }\textbf {\bibinfo {volume} {106}},\
  \bibinfo {pages} {874} (\bibinfo {year} {1957})}\BibitemShut {NoStop}%
\bibitem [{\citenamefont {Ando}\ \emph {et~al.}(1982)\citenamefont {Ando},
  \citenamefont {Fowler},\ and\ \citenamefont {Stern}}]{Ando:1982aa}%
  \BibitemOpen
  \bibfield  {author} {\bibinfo {author} {\bibfnamefont {T.}~\bibnamefont
  {Ando}}, \bibinfo {author} {\bibfnamefont {A.~B.}\ \bibnamefont {Fowler}}, \
  and\ \bibinfo {author} {\bibfnamefont {F.}~\bibnamefont {Stern}},\ }\href
  {\doibase 10.1103/RevModPhys.54.437} {\bibfield  {journal} {\bibinfo
  {journal} {Reviews of Modern Physics}\ }\textbf {\bibinfo {volume} {54}},\
  \bibinfo {pages} {437} (\bibinfo {year} {1982})}\BibitemShut {NoStop}%
\bibitem [{\citenamefont {Stauber}(2014)}]{Stauber:2014aa}%
  \BibitemOpen
  \bibfield  {author} {\bibinfo {author} {\bibfnamefont {T.}~\bibnamefont
  {Stauber}},\ }\href@noop {} {\bibfield  {journal} {\bibinfo  {journal} {J.
  Phys.: Condens. Matter}\ }\textbf {\bibinfo {volume} {26}},\ \bibinfo {pages}
  {123201} (\bibinfo {year} {2014})}\BibitemShut {NoStop}%
\bibitem [{\citenamefont {Stauber}\ \emph
  {et~al.}(2013{\natexlab{a}})\citenamefont {Stauber}, \citenamefont
  {G{\'o}mez-Santos},\ and\ \citenamefont {Brey}}]{Stauber:2013ac}%
  \BibitemOpen
  \bibfield  {author} {\bibinfo {author} {\bibfnamefont {T.}~\bibnamefont
  {Stauber}}, \bibinfo {author} {\bibfnamefont {G.}~\bibnamefont
  {G{\'o}mez-Santos}}, \ and\ \bibinfo {author} {\bibfnamefont
  {L.}~\bibnamefont {Brey}},\ }\href {\doibase 10.1103/PhysRevB.88.205427}
  {\bibfield  {journal} {\bibinfo  {journal} {Physical Review B}\ }\textbf
  {\bibinfo {volume} {88}},\ \bibinfo {pages} {205427} (\bibinfo {year}
  {2013}{\natexlab{a}})}\BibitemShut {NoStop}%
\bibitem [{\citenamefont {Stauber}\ \emph {et~al.}(2017)\citenamefont
  {Stauber}, \citenamefont {G{\'o}mez-Santos},\ and\ \citenamefont
  {Brey}}]{Stauber17}%
  \BibitemOpen
  \bibfield  {author} {\bibinfo {author} {\bibfnamefont {T.}~\bibnamefont
  {Stauber}}, \bibinfo {author} {\bibfnamefont {G.}~\bibnamefont
  {G{\'o}mez-Santos}}, \ and\ \bibinfo {author} {\bibfnamefont
  {L.}~\bibnamefont {Brey}},\ }\href {\doibase 10.1021/acsphotonics.7b00524}
  {\bibfield  {journal} {\bibinfo  {journal} {ACS Photonics}\ }\textbf
  {\bibinfo {volume} {4}},\ \bibinfo {pages} {2978} (\bibinfo {year}
  {2017})}\BibitemShut {NoStop}%
\bibitem [{\citenamefont {Nikitin}(2017)}]{nikitin-book}%
  \BibitemOpen
  \bibfield  {author} {\bibinfo {author} {\bibfnamefont {A.}~\bibnamefont
  {Nikitin}},\ }\href@noop {} {\emph {\bibinfo {title} {Graphene
  Plasmonics,}}},\ \bibinfo {series} {World Scientific Handbook of
  Metamateriasl and Plasmonics}, Vol.~\bibinfo {volume} {1}\ (\bibinfo
  {publisher} {Word Scientific Publishing},\ \bibinfo {year}
  {2017})\BibitemShut {NoStop}%
\bibitem [{\citenamefont {Nikitin}\ \emph {et~al.}(2011)\citenamefont
  {Nikitin}, \citenamefont {Guinea}, \citenamefont {Garc{\'\i}a-Vidal},\ and\
  \citenamefont {Mart{\'\i}n-Moreno}}]{Nikitin:2011aa}%
  \BibitemOpen
  \bibfield  {author} {\bibinfo {author} {\bibfnamefont {A.~Y.}\ \bibnamefont
  {Nikitin}}, \bibinfo {author} {\bibfnamefont {F.}~\bibnamefont {Guinea}},
  \bibinfo {author} {\bibfnamefont {F.~J.}\ \bibnamefont {Garc{\'\i}a-Vidal}},
  \ and\ \bibinfo {author} {\bibfnamefont {L.}~\bibnamefont
  {Mart{\'\i}n-Moreno}},\ }\href {\doibase 10.1103/PhysRevB.84.161407}
  {\bibfield  {journal} {\bibinfo  {journal} {Physical Review B}\ }\textbf
  {\bibinfo {volume} {84}},\ \bibinfo {pages} {161407} (\bibinfo {year}
  {2011})}\BibitemShut {NoStop}%
\bibitem [{\citenamefont {Koppens}\ \emph {et~al.}(2011)\citenamefont
  {Koppens}, \citenamefont {Chang},\ and\ \citenamefont {Garc{\'\i}a~de
  Abajo}}]{Koppens:2011ab}%
  \BibitemOpen
  \bibfield  {author} {\bibinfo {author} {\bibfnamefont {F.~H.~L.}\
  \bibnamefont {Koppens}}, \bibinfo {author} {\bibfnamefont {D.~E.}\
  \bibnamefont {Chang}}, \ and\ \bibinfo {author} {\bibfnamefont {F.~J.}\
  \bibnamefont {Garc{\'\i}a~de Abajo}},\ }\bibfield  {booktitle} {\emph
  {\bibinfo {booktitle} {Nano Letters}},\ }\href {\doibase 10.1021/nl201771h}
  {\bibfield  {journal} {\bibinfo  {journal} {Nano Letters}\ }\textbf {\bibinfo
  {volume} {11}},\ \bibinfo {pages} {3370} (\bibinfo {year}
  {2011})}\BibitemShut {NoStop}%
\bibitem [{\citenamefont {Slipchenko}\ \emph {et~al.}(2013)\citenamefont
  {Slipchenko}, \citenamefont {Nesterov}, \citenamefont {Martin-Moreno},\ and\
  \citenamefont {Nikitin}}]{Slipchenko:2013aa}%
  \BibitemOpen
  \bibfield  {author} {\bibinfo {author} {\bibfnamefont {T.~M.}\ \bibnamefont
  {Slipchenko}}, \bibinfo {author} {\bibfnamefont {M.~L.}\ \bibnamefont
  {Nesterov}}, \bibinfo {author} {\bibfnamefont {L.}~\bibnamefont
  {Martin-Moreno}}, \ and\ \bibinfo {author} {\bibfnamefont {A.~Y.}\
  \bibnamefont {Nikitin}},\ }\href {\doibase 10.1088/2040-8978/15/11/114008}
  {\bibfield  {journal} {\bibinfo  {journal} {Journal of Optics}\ }\textbf
  {\bibinfo {volume} {15}},\ \bibinfo {pages} {114008} (\bibinfo {year}
  {2013})}\BibitemShut {NoStop}%
\bibitem [{\citenamefont {Castro~Neto}\ \emph {et~al.}(2009)\citenamefont
  {Castro~Neto}, \citenamefont {Guinea}, \citenamefont {Peres}, \citenamefont
  {Novoselov},\ and\ \citenamefont {Geim}}]{Guinea_2009}%
  \BibitemOpen
  \bibfield  {author} {\bibinfo {author} {\bibfnamefont {A.~H.}\ \bibnamefont
  {Castro~Neto}}, \bibinfo {author} {\bibfnamefont {F.}~\bibnamefont {Guinea}},
  \bibinfo {author} {\bibfnamefont {N.~M.~R.}\ \bibnamefont {Peres}}, \bibinfo
  {author} {\bibfnamefont {K.~S.}\ \bibnamefont {Novoselov}}, \ and\ \bibinfo
  {author} {\bibfnamefont {A.~K.}\ \bibnamefont {Geim}},\ }\href {\doibase
  10.1103/RevModPhys.81.109} {\bibfield  {journal} {\bibinfo  {journal} {Rev.
  Mod. Phys.}\ }\textbf {\bibinfo {volume} {81}},\ \bibinfo {pages} {109}
  (\bibinfo {year} {2009})}\BibitemShut {NoStop}%
\bibitem [{\citenamefont {M.I.Katsnelson}(2012)}]{Katsnelson-book}%
  \BibitemOpen
  \bibfield  {author} {\bibinfo {author} {\bibnamefont {M.I.Katsnelson}},\
  }\href@noop {} {\emph {\bibinfo {title} {Graphene}}}\ (\bibinfo  {publisher}
  {Cambridge},\ \bibinfo {year} {2012})\BibitemShut {NoStop}%
\bibitem [{\citenamefont {Chen}\ \emph
  {et~al.}(2012{\natexlab{a}})\citenamefont {Chen}, \citenamefont {Badioli},
  \citenamefont {Alonso-Gonz{\'a}lez}, \citenamefont {Thongrattanasiri},
  \citenamefont {Huth}, \citenamefont {Osmond}, \citenamefont {Spasenovi{\'c}},
  \citenamefont {Centeno}, \citenamefont {Pesquera}, \citenamefont {Godignon},
  \citenamefont {Zurutuza~Elorza}, \citenamefont {Camara}, \citenamefont
  {de~Abajo}, \citenamefont {Hillenbrand},\ and\ \citenamefont
  {Koppens}}]{Chen:2012aa}%
  \BibitemOpen
  \bibfield  {author} {\bibinfo {author} {\bibfnamefont {J.}~\bibnamefont
  {Chen}}, \bibinfo {author} {\bibfnamefont {M.}~\bibnamefont {Badioli}},
  \bibinfo {author} {\bibfnamefont {P.}~\bibnamefont {Alonso-Gonz{\'a}lez}},
  \bibinfo {author} {\bibfnamefont {S.}~\bibnamefont {Thongrattanasiri}},
  \bibinfo {author} {\bibfnamefont {F.}~\bibnamefont {Huth}}, \bibinfo {author}
  {\bibfnamefont {J.}~\bibnamefont {Osmond}}, \bibinfo {author} {\bibfnamefont
  {M.}~\bibnamefont {Spasenovi{\'c}}}, \bibinfo {author} {\bibfnamefont
  {A.}~\bibnamefont {Centeno}}, \bibinfo {author} {\bibfnamefont
  {A.}~\bibnamefont {Pesquera}}, \bibinfo {author} {\bibfnamefont
  {P.}~\bibnamefont {Godignon}}, \bibinfo {author} {\bibfnamefont
  {A.}~\bibnamefont {Zurutuza~Elorza}}, \bibinfo {author} {\bibfnamefont
  {N.}~\bibnamefont {Camara}}, \bibinfo {author} {\bibfnamefont {F.~J.~G.}\
  \bibnamefont {de~Abajo}}, \bibinfo {author} {\bibfnamefont {R.}~\bibnamefont
  {Hillenbrand}}, \ and\ \bibinfo {author} {\bibfnamefont {F.~H.~L.}\
  \bibnamefont {Koppens}},\ }\href {\doibase 10.1038/nature11254} {\bibfield
  {journal} {\bibinfo  {journal} {Nature}\ }\textbf {\bibinfo {volume} {487}},\
  \bibinfo {pages} {77} (\bibinfo {year} {2012}{\natexlab{a}})}\BibitemShut
  {NoStop}%
\bibitem [{\citenamefont {Fei}\ \emph {et~al.}(2012{\natexlab{a}})\citenamefont
  {Fei}, \citenamefont {Rodin}, \citenamefont {Andreev}, \citenamefont {Bao},
  \citenamefont {McLeod}, \citenamefont {Wagner}, \citenamefont {Zhang},
  \citenamefont {Zhao}, \citenamefont {Thiemens}, \citenamefont {Dominguez},
  \citenamefont {Fogler}, \citenamefont {Neto}, \citenamefont {Lau},
  \citenamefont {Keilmann},\ and\ \citenamefont {Basov}}]{Fei:2012aa}%
  \BibitemOpen
  \bibfield  {author} {\bibinfo {author} {\bibfnamefont {Z.}~\bibnamefont
  {Fei}}, \bibinfo {author} {\bibfnamefont {A.~S.}\ \bibnamefont {Rodin}},
  \bibinfo {author} {\bibfnamefont {G.~O.}\ \bibnamefont {Andreev}}, \bibinfo
  {author} {\bibfnamefont {W.}~\bibnamefont {Bao}}, \bibinfo {author}
  {\bibfnamefont {A.~S.}\ \bibnamefont {McLeod}}, \bibinfo {author}
  {\bibfnamefont {M.}~\bibnamefont {Wagner}}, \bibinfo {author} {\bibfnamefont
  {L.~M.}\ \bibnamefont {Zhang}}, \bibinfo {author} {\bibfnamefont
  {Z.}~\bibnamefont {Zhao}}, \bibinfo {author} {\bibfnamefont {M.}~\bibnamefont
  {Thiemens}}, \bibinfo {author} {\bibfnamefont {G.}~\bibnamefont {Dominguez}},
  \bibinfo {author} {\bibfnamefont {M.~M.}\ \bibnamefont {Fogler}}, \bibinfo
  {author} {\bibfnamefont {A.~H.~C.}\ \bibnamefont {Neto}}, \bibinfo {author}
  {\bibfnamefont {C.~N.}\ \bibnamefont {Lau}}, \bibinfo {author} {\bibfnamefont
  {F.}~\bibnamefont {Keilmann}}, \ and\ \bibinfo {author} {\bibfnamefont
  {D.~N.}\ \bibnamefont {Basov}},\ }\href {\doibase 10.1038/nature11253}
  {\bibfield  {journal} {\bibinfo  {journal} {Nature}\ }\textbf {\bibinfo
  {volume} {487}},\ \bibinfo {pages} {82} (\bibinfo {year}
  {2012}{\natexlab{a}})}\BibitemShut {NoStop}%
\bibitem [{\citenamefont {Fei}\ \emph {et~al.}(2013)\citenamefont {Fei},
  \citenamefont {Rodin}, \citenamefont {Gannett}, \citenamefont {Dai},
  \citenamefont {Regan}, \citenamefont {Wagner}, \citenamefont {Liu},
  \citenamefont {McLeod}, \citenamefont {Dominguez}, \citenamefont {Thiemens},
  \citenamefont {Castro~Neto}, \citenamefont {Keilmann}, \citenamefont {Zettl},
  \citenamefont {Hillenbrand}, \citenamefont {Fogler},\ and\ \citenamefont
  {Basov}}]{Fei:2013aa}%
  \BibitemOpen
  \bibfield  {author} {\bibinfo {author} {\bibfnamefont {Z.}~\bibnamefont
  {Fei}}, \bibinfo {author} {\bibfnamefont {A.~S.}\ \bibnamefont {Rodin}},
  \bibinfo {author} {\bibfnamefont {W.}~\bibnamefont {Gannett}}, \bibinfo
  {author} {\bibfnamefont {S.}~\bibnamefont {Dai}}, \bibinfo {author}
  {\bibfnamefont {W.}~\bibnamefont {Regan}}, \bibinfo {author} {\bibfnamefont
  {M.}~\bibnamefont {Wagner}}, \bibinfo {author} {\bibfnamefont {M.~K.}\
  \bibnamefont {Liu}}, \bibinfo {author} {\bibfnamefont {A.~S.}\ \bibnamefont
  {McLeod}}, \bibinfo {author} {\bibfnamefont {G.}~\bibnamefont {Dominguez}},
  \bibinfo {author} {\bibfnamefont {M.}~\bibnamefont {Thiemens}}, \bibinfo
  {author} {\bibfnamefont {A.~H.}\ \bibnamefont {Castro~Neto}}, \bibinfo
  {author} {\bibfnamefont {F.}~\bibnamefont {Keilmann}}, \bibinfo {author}
  {\bibfnamefont {A.}~\bibnamefont {Zettl}}, \bibinfo {author} {\bibfnamefont
  {R.}~\bibnamefont {Hillenbrand}}, \bibinfo {author} {\bibfnamefont {M.~M.}\
  \bibnamefont {Fogler}}, \ and\ \bibinfo {author} {\bibfnamefont {D.~N.}\
  \bibnamefont {Basov}},\ }\href {\doibase 10.1038/nnano.2013.197} {\bibfield
  {journal} {\bibinfo  {journal} {Nature Nanotechnology}\ }\textbf {\bibinfo
  {volume} {8}},\ \bibinfo {pages} {821} (\bibinfo {year} {2013})}\BibitemShut
  {NoStop}%
\bibitem [{\citenamefont {Ju}\ \emph {et~al.}(2011)\citenamefont {Ju},
  \citenamefont {Geng}, \citenamefont {Horng}, \citenamefont {Girit},
  \citenamefont {Martin}, \citenamefont {Hao}, \citenamefont {Bechtel},
  \citenamefont {Liang}, \citenamefont {Zettl}, \citenamefont {Shen},\ and\
  \citenamefont {Wang}}]{Ju11}%
  \BibitemOpen
  \bibfield  {author} {\bibinfo {author} {\bibfnamefont {L.}~\bibnamefont
  {Ju}}, \bibinfo {author} {\bibfnamefont {B.}~\bibnamefont {Geng}}, \bibinfo
  {author} {\bibfnamefont {J.}~\bibnamefont {Horng}}, \bibinfo {author}
  {\bibfnamefont {C.}~\bibnamefont {Girit}}, \bibinfo {author} {\bibfnamefont
  {M.}~\bibnamefont {Martin}}, \bibinfo {author} {\bibfnamefont
  {Z.}~\bibnamefont {Hao}}, \bibinfo {author} {\bibfnamefont {H.~A.}\
  \bibnamefont {Bechtel}}, \bibinfo {author} {\bibfnamefont {X.}~\bibnamefont
  {Liang}}, \bibinfo {author} {\bibfnamefont {A.}~\bibnamefont {Zettl}},
  \bibinfo {author} {\bibfnamefont {Y.~R.}\ \bibnamefont {Shen}}, \ and\
  \bibinfo {author} {\bibfnamefont {F.}~\bibnamefont {Wang}},\ }\href {\doibase
  10.1038/nnano.2011.146} {\bibfield  {journal} {\bibinfo  {journal} {Nature
  Nanotechnology}\ }\textbf {\bibinfo {volume} {6}},\ \bibinfo {pages} {630}
  (\bibinfo {year} {2011})}\BibitemShut {NoStop}%
\bibitem [{\citenamefont {Bonaccorso}\ \emph {et~al.}(2010)\citenamefont
  {Bonaccorso}, \citenamefont {Sun}, \citenamefont {Hasan},\ and\ \citenamefont
  {Ferrari}}]{Bonaccorso:2010aa}%
  \BibitemOpen
  \bibfield  {author} {\bibinfo {author} {\bibfnamefont {F.}~\bibnamefont
  {Bonaccorso}}, \bibinfo {author} {\bibfnamefont {Z.}~\bibnamefont {Sun}},
  \bibinfo {author} {\bibfnamefont {T.}~\bibnamefont {Hasan}}, \ and\ \bibinfo
  {author} {\bibfnamefont {A.~C.}\ \bibnamefont {Ferrari}},\ }\href
  {https://doi.org/10.1038/nphoton.2010.186} {\bibfield  {journal} {\bibinfo
  {journal} {Nature Photonics}\ }\textbf {\bibinfo {volume} {4}},\ \bibinfo
  {pages} {611 EP } (\bibinfo {year} {2010})}\BibitemShut {NoStop}%
\bibitem [{\citenamefont {Farmer}\ \emph {et~al.}(2015)\citenamefont {Farmer},
  \citenamefont {Rodrigo}, \citenamefont {Low},\ and\ \citenamefont
  {Avouris}}]{Farmer:2015aa}%
  \BibitemOpen
  \bibfield  {author} {\bibinfo {author} {\bibfnamefont {D.~B.}\ \bibnamefont
  {Farmer}}, \bibinfo {author} {\bibfnamefont {D.}~\bibnamefont {Rodrigo}},
  \bibinfo {author} {\bibfnamefont {T.}~\bibnamefont {Low}}, \ and\ \bibinfo
  {author} {\bibfnamefont {P.}~\bibnamefont {Avouris}},\ }\href {\doibase
  10.1021/acs.nanolett.5b00148} {\bibfield  {journal} {\bibinfo  {journal}
  {Nano Letters}\ }\textbf {\bibinfo {volume} {15}},\ \bibinfo {pages} {2582}
  (\bibinfo {year} {2015})}\BibitemShut {NoStop}%
\bibitem [{\citenamefont {Ni}\ \emph {et~al.}(2015)\citenamefont {Ni},
  \citenamefont {Wang}, \citenamefont {Wu}, \citenamefont {Fei}, \citenamefont
  {Goldflam}, \citenamefont {Keilmann}, \citenamefont {Ozyilmaz}, \citenamefont
  {Castro~Neto}, \citenamefont {Xie}, \citenamefont {Fogler},\ and\
  \citenamefont {Basov}}]{Ni15}%
  \BibitemOpen
  \bibfield  {author} {\bibinfo {author} {\bibfnamefont {G.~X.}\ \bibnamefont
  {Ni}}, \bibinfo {author} {\bibfnamefont {H.}~\bibnamefont {Wang}}, \bibinfo
  {author} {\bibfnamefont {J.~S.}\ \bibnamefont {Wu}}, \bibinfo {author}
  {\bibfnamefont {Z.}~\bibnamefont {Fei}}, \bibinfo {author} {\bibfnamefont
  {M.~D.}\ \bibnamefont {Goldflam}}, \bibinfo {author} {\bibfnamefont
  {F.}~\bibnamefont {Keilmann}}, \bibinfo {author} {\bibfnamefont
  {B.}~\bibnamefont {Ozyilmaz}}, \bibinfo {author} {\bibfnamefont {A.~H.}\
  \bibnamefont {Castro~Neto}}, \bibinfo {author} {\bibfnamefont {X.~M.}\
  \bibnamefont {Xie}}, \bibinfo {author} {\bibfnamefont {M.~M.}\ \bibnamefont
  {Fogler}}, \ and\ \bibinfo {author} {\bibfnamefont {D.~N.}\ \bibnamefont
  {Basov}},\ }\href {http://dx.doi.org/10.1038/nmat4425} {\bibfield  {journal}
  {\bibinfo  {journal} {Nat Mater}\ }\textbf {\bibinfo {volume} {14}},\
  \bibinfo {pages} {1217} (\bibinfo {year} {2015})}\BibitemShut {NoStop}%
\bibitem [{\citenamefont {Liu}\ \emph {et~al.}(2015)\citenamefont {Liu},
  \citenamefont {Valmorra}, \citenamefont {Maissen},\ and\ \citenamefont
  {Faist}}]{Liu:2015aa}%
  \BibitemOpen
  \bibfield  {author} {\bibinfo {author} {\bibfnamefont {P.~Q.}\ \bibnamefont
  {Liu}}, \bibinfo {author} {\bibfnamefont {F.}~\bibnamefont {Valmorra}},
  \bibinfo {author} {\bibfnamefont {C.}~\bibnamefont {Maissen}}, \ and\
  \bibinfo {author} {\bibfnamefont {J.}~\bibnamefont {Faist}},\ }\bibfield
  {booktitle} {\emph {\bibinfo {booktitle} {Optica}},\ }\href {\doibase
  10.1364/OPTICA.2.000135} {\bibfield  {journal} {\bibinfo  {journal} {Optica}\
  }\textbf {\bibinfo {volume} {2}},\ \bibinfo {pages} {135} (\bibinfo {year}
  {2015})}\BibitemShut {NoStop}%
\bibitem [{\citenamefont {Alcaraz~Iranzo}\ \emph {et~al.}(2018)\citenamefont
  {Alcaraz~Iranzo}, \citenamefont {Nanot}, \citenamefont {Dias}, \citenamefont
  {Epstein}, \citenamefont {Peng}, \citenamefont {Efetov}, \citenamefont
  {Lundeberg}, \citenamefont {Parret}, \citenamefont {Osmond}, \citenamefont
  {Hong}, \citenamefont {Kong}, \citenamefont {Englund}, \citenamefont
  {Peres},\ and\ \citenamefont {Koppens}}]{Alcaraz18}%
  \BibitemOpen
  \bibfield  {author} {\bibinfo {author} {\bibfnamefont {D.}~\bibnamefont
  {Alcaraz~Iranzo}}, \bibinfo {author} {\bibfnamefont {S.}~\bibnamefont
  {Nanot}}, \bibinfo {author} {\bibfnamefont {E.~J.~C.}\ \bibnamefont {Dias}},
  \bibinfo {author} {\bibfnamefont {I.}~\bibnamefont {Epstein}}, \bibinfo
  {author} {\bibfnamefont {C.}~\bibnamefont {Peng}}, \bibinfo {author}
  {\bibfnamefont {D.~K.}\ \bibnamefont {Efetov}}, \bibinfo {author}
  {\bibfnamefont {M.~B.}\ \bibnamefont {Lundeberg}}, \bibinfo {author}
  {\bibfnamefont {R.}~\bibnamefont {Parret}}, \bibinfo {author} {\bibfnamefont
  {J.}~\bibnamefont {Osmond}}, \bibinfo {author} {\bibfnamefont {J.-Y.}\
  \bibnamefont {Hong}}, \bibinfo {author} {\bibfnamefont {J.}~\bibnamefont
  {Kong}}, \bibinfo {author} {\bibfnamefont {D.~R.}\ \bibnamefont {Englund}},
  \bibinfo {author} {\bibfnamefont {N.~M.~R.}\ \bibnamefont {Peres}}, \ and\
  \bibinfo {author} {\bibfnamefont {F.~H.~L.}\ \bibnamefont {Koppens}},\ }\href
  {\doibase 10.1126/science.aar8438} {\bibfield  {journal} {\bibinfo  {journal}
  {Science}\ }\textbf {\bibinfo {volume} {360}},\ \bibinfo {pages} {291}
  (\bibinfo {year} {2018})}\BibitemShut {NoStop}%
\bibitem [{\citenamefont {Gopalan}\ \emph {et~al.}(2018)\citenamefont
  {Gopalan}, \citenamefont {Paulillo}, \citenamefont {Mackenzie}, \citenamefont
  {Rodrigo}, \citenamefont {Bareza}, \citenamefont {Whelan}, \citenamefont
  {Shivayogimath},\ and\ \citenamefont {Pruneri}}]{Gopalan:2018aa}%
  \BibitemOpen
  \bibfield  {author} {\bibinfo {author} {\bibfnamefont {K.~K.}\ \bibnamefont
  {Gopalan}}, \bibinfo {author} {\bibfnamefont {B.}~\bibnamefont {Paulillo}},
  \bibinfo {author} {\bibfnamefont {D.~M.~A.}\ \bibnamefont {Mackenzie}},
  \bibinfo {author} {\bibfnamefont {D.}~\bibnamefont {Rodrigo}}, \bibinfo
  {author} {\bibfnamefont {N.}~\bibnamefont {Bareza}}, \bibinfo {author}
  {\bibfnamefont {P.~R.}\ \bibnamefont {Whelan}}, \bibinfo {author}
  {\bibfnamefont {A.}~\bibnamefont {Shivayogimath}}, \ and\ \bibinfo {author}
  {\bibfnamefont {V.}~\bibnamefont {Pruneri}},\ }\bibfield  {booktitle} {\emph
  {\bibinfo {booktitle} {Nano Letters}},\ }\href {\doibase
  10.1021/acs.nanolett.8b02613} {\bibfield  {journal} {\bibinfo  {journal}
  {Nano Letters}\ }\textbf {\bibinfo {volume} {18}},\ \bibinfo {pages} {5913}
  (\bibinfo {year} {2018})}\BibitemShut {NoStop}%
\bibitem [{\citenamefont {Nikitin}\ \emph {et~al.}(2012)\citenamefont
  {Nikitin}, \citenamefont {Guinea}, \citenamefont {Garcia-Vidal},\ and\
  \citenamefont {Martin-Moreno}}]{Nikitin:2012aa}%
  \BibitemOpen
  \bibfield  {author} {\bibinfo {author} {\bibfnamefont {A.~Y.}\ \bibnamefont
  {Nikitin}}, \bibinfo {author} {\bibfnamefont {F.}~\bibnamefont {Guinea}},
  \bibinfo {author} {\bibfnamefont {F.~J.}\ \bibnamefont {Garcia-Vidal}}, \
  and\ \bibinfo {author} {\bibfnamefont {L.}~\bibnamefont {Martin-Moreno}},\
  }\href {\doibase 10.1103/PhysRevB.85.081405} {\bibfield  {journal} {\bibinfo
  {journal} {Physical Review B}\ }\textbf {\bibinfo {volume} {85}},\ \bibinfo
  {pages} {081405} (\bibinfo {year} {2012})}\BibitemShut {NoStop}%
\bibitem [{\citenamefont {Vacacela~Gomez}\ \emph {et~al.}(2016)\citenamefont
  {Vacacela~Gomez}, \citenamefont {Pisarra}, \citenamefont {Gravina},
  \citenamefont {Pitarke},\ and\ \citenamefont
  {Sindona}}]{Vacacela-Gomez:2016aa}%
  \BibitemOpen
  \bibfield  {author} {\bibinfo {author} {\bibfnamefont {C.}~\bibnamefont
  {Vacacela~Gomez}}, \bibinfo {author} {\bibfnamefont {M.}~\bibnamefont
  {Pisarra}}, \bibinfo {author} {\bibfnamefont {M.}~\bibnamefont {Gravina}},
  \bibinfo {author} {\bibfnamefont {J.~M.}\ \bibnamefont {Pitarke}}, \ and\
  \bibinfo {author} {\bibfnamefont {A.}~\bibnamefont {Sindona}},\ }\href
  {\doibase 10.1103/PhysRevLett.117.116801} {\bibfield  {journal} {\bibinfo
  {journal} {Physical Review Letters}\ }\textbf {\bibinfo {volume} {117}},\
  \bibinfo {pages} {116801} (\bibinfo {year} {2016})}\BibitemShut {NoStop}%
\bibitem [{\citenamefont {Peres}\ \emph {et~al.}(2012)\citenamefont {Peres},
  \citenamefont {Ferreira}, \citenamefont {Bludov},\ and\ \citenamefont
  {Vasilevskiy}}]{Peres:2012aa}%
  \BibitemOpen
  \bibfield  {author} {\bibinfo {author} {\bibfnamefont {N.~M.~R.}\
  \bibnamefont {Peres}}, \bibinfo {author} {\bibfnamefont {A.}~\bibnamefont
  {Ferreira}}, \bibinfo {author} {\bibfnamefont {Y.~V.}\ \bibnamefont
  {Bludov}}, \ and\ \bibinfo {author} {\bibfnamefont {M.~I.}\ \bibnamefont
  {Vasilevskiy}},\ }\href {\doibase 10.1088/0953-8984/24/24/245303} {\bibfield
  {journal} {\bibinfo  {journal} {Journal of Physics: Condensed Matter}\
  }\textbf {\bibinfo {volume} {24}},\ \bibinfo {pages} {245303} (\bibinfo
  {year} {2012})}\BibitemShut {NoStop}%
\bibitem [{\citenamefont {Sunku}\ \emph {et~al.}(2018)\citenamefont {Sunku},
  \citenamefont {Ni}, \citenamefont {Jiang}, \citenamefont {Yoo}, \citenamefont
  {Sternbach}, \citenamefont {McLeod}, \citenamefont {Stauber}, \citenamefont
  {Xiong}, \citenamefont {Taniguchi}, \citenamefont {Watanabe}, \citenamefont
  {Kim}, \citenamefont {Fogler},\ and\ \citenamefont {Basov}}]{Sunku:2018aa}%
  \BibitemOpen
  \bibfield  {author} {\bibinfo {author} {\bibfnamefont {S.~S.}\ \bibnamefont
  {Sunku}}, \bibinfo {author} {\bibfnamefont {G.~X.}\ \bibnamefont {Ni}},
  \bibinfo {author} {\bibfnamefont {B.~Y.}\ \bibnamefont {Jiang}}, \bibinfo
  {author} {\bibfnamefont {H.}~\bibnamefont {Yoo}}, \bibinfo {author}
  {\bibfnamefont {A.}~\bibnamefont {Sternbach}}, \bibinfo {author}
  {\bibfnamefont {A.~S.}\ \bibnamefont {McLeod}}, \bibinfo {author}
  {\bibfnamefont {T.}~\bibnamefont {Stauber}}, \bibinfo {author} {\bibfnamefont
  {L.}~\bibnamefont {Xiong}}, \bibinfo {author} {\bibfnamefont
  {T.}~\bibnamefont {Taniguchi}}, \bibinfo {author} {\bibfnamefont
  {K.}~\bibnamefont {Watanabe}}, \bibinfo {author} {\bibfnamefont
  {P.}~\bibnamefont {Kim}}, \bibinfo {author} {\bibfnamefont {M.~M.}\
  \bibnamefont {Fogler}}, \ and\ \bibinfo {author} {\bibfnamefont {D.~N.}\
  \bibnamefont {Basov}},\ }\href {\doibase 10.1126/science.aau5144} {\bibfield
  {journal} {\bibinfo  {journal} {Science}\ }\textbf {\bibinfo {volume}
  {362}},\ \bibinfo {pages} {1153} (\bibinfo {year} {2018})}\BibitemShut
  {NoStop}%
\bibitem [{\citenamefont {Brey}\ \emph
  {et~al.}(2020{\natexlab{a}})\citenamefont {Brey}, \citenamefont {Stauber},
  \citenamefont {Mart{\'\i}n-Moreno},\ and\ \citenamefont
  {G{\'o}mez-Santos}}]{Brey:2020aa}%
  \BibitemOpen
  \bibfield  {author} {\bibinfo {author} {\bibfnamefont {L.}~\bibnamefont
  {Brey}}, \bibinfo {author} {\bibfnamefont {T.}~\bibnamefont {Stauber}},
  \bibinfo {author} {\bibfnamefont {L.}~\bibnamefont {Mart{\'\i}n-Moreno}}, \
  and\ \bibinfo {author} {\bibfnamefont {G.}~\bibnamefont {G{\'o}mez-Santos}},\
  }\href {\doibase 10.1103/PhysRevLett.124.257401} {\bibfield  {journal}
  {\bibinfo  {journal} {Physical Review Letters}\ }\textbf {\bibinfo {volume}
  {124}},\ \bibinfo {pages} {257401} (\bibinfo {year}
  {2020}{\natexlab{a}})}\BibitemShut {NoStop}%
\bibitem [{\citenamefont {Brey}\ \emph
  {et~al.}(2020{\natexlab{b}})\citenamefont {Brey}, \citenamefont {Stauber},
  \citenamefont {Slipchenko},\ and\ \citenamefont
  {Mart{\'\i}n-Moreno}}]{Brey:2020ab}%
  \BibitemOpen
  \bibfield  {author} {\bibinfo {author} {\bibfnamefont {L.}~\bibnamefont
  {Brey}}, \bibinfo {author} {\bibfnamefont {T.}~\bibnamefont {Stauber}},
  \bibinfo {author} {\bibfnamefont {T.}~\bibnamefont {Slipchenko}}, \ and\
  \bibinfo {author} {\bibfnamefont {L.}~\bibnamefont {Mart{\'\i}n-Moreno}},\
  }\href {\doibase 10.1103/PhysRevLett.125.256804} {\bibfield  {journal}
  {\bibinfo  {journal} {Physical Review Letters}\ }\textbf {\bibinfo {volume}
  {125}},\ \bibinfo {pages} {256804} (\bibinfo {year}
  {2020}{\natexlab{b}})}\BibitemShut {NoStop}%
\bibitem [{\citenamefont {Stauber}\ \emph
  {et~al.}(2013{\natexlab{b}})\citenamefont {Stauber}, \citenamefont
  {San-Jose},\ and\ \citenamefont {Brey}}]{Stauber:2013aa}%
  \BibitemOpen
  \bibfield  {author} {\bibinfo {author} {\bibfnamefont {T.}~\bibnamefont
  {Stauber}}, \bibinfo {author} {\bibfnamefont {P.}~\bibnamefont {San-Jose}}, \
  and\ \bibinfo {author} {\bibfnamefont {L.}~\bibnamefont {Brey}},\ }\href
  {\doibase 10.1088/1367-2630/15/11/113050} {\bibfield  {journal} {\bibinfo
  {journal} {New Journal of Physics}\ }\textbf {\bibinfo {volume} {15}},\
  \bibinfo {pages} {113050} (\bibinfo {year} {2013}{\natexlab{b}})}\BibitemShut
  {NoStop}%
\bibitem [{\citenamefont {G.~Gilbert}(2010)}]{Grynberg-book}%
  \BibitemOpen
  \bibfield  {author} {\bibinfo {author} {\bibfnamefont {C.~F.}\ \bibnamefont
  {G.~Gilbert}, \bibfnamefont {A.~Aspect}},\ }\href@noop {} {\emph {\bibinfo
  {title} {Introduction to Quantum Optics}}}\ (\bibinfo  {publisher} {Cambridge
  University Press},\ \bibinfo {year} {2010})\BibitemShut {NoStop}%
\bibitem [{\citenamefont {Vogel}\ and\ \citenamefont
  {Welsch}(2006)}]{Vogel-Welsch-Book}%
  \BibitemOpen
  \bibfield  {author} {\bibinfo {author} {\bibfnamefont {W.}~\bibnamefont
  {Vogel}}\ and\ \bibinfo {author} {\bibfnamefont {D.-G.}\ \bibnamefont
  {Welsch}},\ }\href@noop {} {\emph {\bibinfo {title} {Quantum Optics}}}\
  (\bibinfo  {publisher} {Wiley},\ \bibinfo {year} {2006})\BibitemShut
  {NoStop}%
\bibitem [{\citenamefont {T{\"o}rm{\"a}}\ and\ \citenamefont
  {Barnes}(2015)}]{Torma:2015aa}%
  \BibitemOpen
  \bibfield  {author} {\bibinfo {author} {\bibfnamefont {P.}~\bibnamefont
  {T{\"o}rm{\"a}}}\ and\ \bibinfo {author} {\bibfnamefont {W.~L.}\ \bibnamefont
  {Barnes}},\ }\bibfield  {booktitle} {\emph {\bibinfo {booktitle} {Reports on
  Progress in Physics}},\ }\href {\doibase 10.1088/0034-4885/78/1/013901}
  {\bibfield  {journal} {\bibinfo  {journal} {Reports on Progress in Physics}\
  }\textbf {\bibinfo {volume} {78}},\ \bibinfo {pages} {013901} (\bibinfo
  {year} {2015})}\BibitemShut {NoStop}%
\bibitem [{\citenamefont {Ciuti}\ \emph {et~al.}(2005)\citenamefont {Ciuti},
  \citenamefont {Bastard},\ and\ \citenamefont {Carusotto}}]{Ciuti:2005aa}%
  \BibitemOpen
  \bibfield  {author} {\bibinfo {author} {\bibfnamefont {C.}~\bibnamefont
  {Ciuti}}, \bibinfo {author} {\bibfnamefont {G.}~\bibnamefont {Bastard}}, \
  and\ \bibinfo {author} {\bibfnamefont {I.}~\bibnamefont {Carusotto}},\ }\href
  {\doibase 10.1103/PhysRevB.72.115303} {\bibfield  {journal} {\bibinfo
  {journal} {Physical Review B}\ }\textbf {\bibinfo {volume} {72}},\ \bibinfo
  {pages} {115303} (\bibinfo {year} {2005})}\BibitemShut {NoStop}%
\bibitem [{\citenamefont {Berthel}\ \emph {et~al.}(2016)\citenamefont
  {Berthel}, \citenamefont {Huant},\ and\ \citenamefont
  {Drezet}}]{Berthel:2016aa}%
  \BibitemOpen
  \bibfield  {author} {\bibinfo {author} {\bibfnamefont {M.}~\bibnamefont
  {Berthel}}, \bibinfo {author} {\bibfnamefont {S.}~\bibnamefont {Huant}}, \
  and\ \bibinfo {author} {\bibfnamefont {A.}~\bibnamefont {Drezet}},\
  }\bibfield  {booktitle} {\emph {\bibinfo {booktitle} {Optics Letters}},\
  }\href {\doibase 10.1364/OL.41.000037} {\bibfield  {journal} {\bibinfo
  {journal} {Optics Letters}\ }\textbf {\bibinfo {volume} {41}},\ \bibinfo
  {pages} {37} (\bibinfo {year} {2016})}\BibitemShut {NoStop}%
\bibitem [{\citenamefont {Sun}\ \emph {et~al.}(2022)\citenamefont {Sun},
  \citenamefont {Basov},\ and\ \citenamefont {Fogler}}]{Sun:2022aa}%
  \BibitemOpen
  \bibfield  {author} {\bibinfo {author} {\bibfnamefont {Z.}~\bibnamefont
  {Sun}}, \bibinfo {author} {\bibfnamefont {D.~N.}\ \bibnamefont {Basov}}, \
  and\ \bibinfo {author} {\bibfnamefont {M.~M.}\ \bibnamefont {Fogler}},\
  }\href {\doibase 10.1103/PhysRevResearch.4.023208} {\bibfield  {journal}
  {\bibinfo  {journal} {Physical Review Research}\ }\textbf {\bibinfo {volume}
  {4}},\ \bibinfo {pages} {023208} (\bibinfo {year} {2022})}\BibitemShut
  {NoStop}%
\bibitem [{\citenamefont {Chang}\ \emph {et~al.}(2006)\citenamefont {Chang},
  \citenamefont {S{\o}rensen}, \citenamefont {Hemmer},\ and\ \citenamefont
  {Lukin}}]{Chang:2006aa}%
  \BibitemOpen
  \bibfield  {author} {\bibinfo {author} {\bibfnamefont {D.~E.}\ \bibnamefont
  {Chang}}, \bibinfo {author} {\bibfnamefont {A.~S.}\ \bibnamefont
  {S{\o}rensen}}, \bibinfo {author} {\bibfnamefont {P.~R.}\ \bibnamefont
  {Hemmer}}, \ and\ \bibinfo {author} {\bibfnamefont {M.~D.}\ \bibnamefont
  {Lukin}},\ }\href {\doibase 10.1103/PhysRevLett.97.053002} {\bibfield
  {journal} {\bibinfo  {journal} {Physical Review Letters}\ }\textbf {\bibinfo
  {volume} {97}},\ \bibinfo {pages} {053002} (\bibinfo {year}
  {2006})}\BibitemShut {NoStop}%
\bibitem [{\citenamefont {Gonz{\'a}lez-Tudela}\ \emph
  {et~al.}(2013)\citenamefont {Gonz{\'a}lez-Tudela}, \citenamefont {Huidobro},
  \citenamefont {Mart{\'\i}n-Moreno}, \citenamefont {Tejedor},\ and\
  \citenamefont {Garc{\'\i}a-Vidal}}]{Gonzalez-Tudela:2013aa}%
  \BibitemOpen
  \bibfield  {author} {\bibinfo {author} {\bibfnamefont {A.}~\bibnamefont
  {Gonz{\'a}lez-Tudela}}, \bibinfo {author} {\bibfnamefont {P.~A.}\
  \bibnamefont {Huidobro}}, \bibinfo {author} {\bibfnamefont {L.}~\bibnamefont
  {Mart{\'\i}n-Moreno}}, \bibinfo {author} {\bibfnamefont {C.}~\bibnamefont
  {Tejedor}}, \ and\ \bibinfo {author} {\bibfnamefont {F.~J.}\ \bibnamefont
  {Garc{\'\i}a-Vidal}},\ }\href {\doibase 10.1103/PhysRevLett.110.126801}
  {\bibfield  {journal} {\bibinfo  {journal} {Physical Review Letters}\
  }\textbf {\bibinfo {volume} {110}},\ \bibinfo {pages} {126801} (\bibinfo
  {year} {2013})}\BibitemShut {NoStop}%
\bibitem [{\citenamefont {Gonzalez-Tudela}\ \emph {et~al.}(2011)\citenamefont
  {Gonzalez-Tudela}, \citenamefont {Martin-Cano}, \citenamefont {Moreno},
  \citenamefont {Martin-Moreno}, \citenamefont {Tejedor},\ and\ \citenamefont
  {Garcia-Vidal}}]{Gonzalez-Tudela:2011aa}%
  \BibitemOpen
  \bibfield  {author} {\bibinfo {author} {\bibfnamefont {A.}~\bibnamefont
  {Gonzalez-Tudela}}, \bibinfo {author} {\bibfnamefont {D.}~\bibnamefont
  {Martin-Cano}}, \bibinfo {author} {\bibfnamefont {E.}~\bibnamefont {Moreno}},
  \bibinfo {author} {\bibfnamefont {L.}~\bibnamefont {Martin-Moreno}}, \bibinfo
  {author} {\bibfnamefont {C.}~\bibnamefont {Tejedor}}, \ and\ \bibinfo
  {author} {\bibfnamefont {F.~J.}\ \bibnamefont {Garcia-Vidal}},\ }\href
  {\doibase 10.1103/PhysRevLett.106.020501} {\bibfield  {journal} {\bibinfo
  {journal} {Physical Review Letters}\ }\textbf {\bibinfo {volume} {106}},\
  \bibinfo {pages} {020501} (\bibinfo {year} {2011})}\BibitemShut {NoStop}%
\bibitem [{\citenamefont {Ponomarenko}\ \emph {et~al.}(2011)\citenamefont
  {Ponomarenko}, \citenamefont {Geim}, \citenamefont {Zhukov}, \citenamefont
  {Jalil}, \citenamefont {Morozov}, \citenamefont {Novoselov}, \citenamefont
  {Grigorieva}, \citenamefont {Hill}, \citenamefont {Cheianov}, \citenamefont
  {Fal'ko}, \citenamefont {Watanabe}, \citenamefont {Taniguchi},\ and\
  \citenamefont {Gorbachev}}]{Ponomarenko:2011aa}%
  \BibitemOpen
  \bibfield  {author} {\bibinfo {author} {\bibfnamefont {L.~A.}\ \bibnamefont
  {Ponomarenko}}, \bibinfo {author} {\bibfnamefont {A.~K.}\ \bibnamefont
  {Geim}}, \bibinfo {author} {\bibfnamefont {A.~A.}\ \bibnamefont {Zhukov}},
  \bibinfo {author} {\bibfnamefont {R.}~\bibnamefont {Jalil}}, \bibinfo
  {author} {\bibfnamefont {S.~V.}\ \bibnamefont {Morozov}}, \bibinfo {author}
  {\bibfnamefont {K.~S.}\ \bibnamefont {Novoselov}}, \bibinfo {author}
  {\bibfnamefont {I.~V.}\ \bibnamefont {Grigorieva}}, \bibinfo {author}
  {\bibfnamefont {E.~H.}\ \bibnamefont {Hill}}, \bibinfo {author}
  {\bibfnamefont {V.~V.}\ \bibnamefont {Cheianov}}, \bibinfo {author}
  {\bibfnamefont {V.~I.}\ \bibnamefont {Fal'ko}}, \bibinfo {author}
  {\bibfnamefont {K.}~\bibnamefont {Watanabe}}, \bibinfo {author}
  {\bibfnamefont {T.}~\bibnamefont {Taniguchi}}, \ and\ \bibinfo {author}
  {\bibfnamefont {R.~V.}\ \bibnamefont {Gorbachev}},\ }\href {\doibase
  10.1038/nphys2114} {\bibfield  {journal} {\bibinfo  {journal} {Nature
  Physics}\ }\textbf {\bibinfo {volume} {7}},\ \bibinfo {pages} {958} (\bibinfo
  {year} {2011})}\BibitemShut {NoStop}%
\bibitem [{\citenamefont {Stauber}\ and\ \citenamefont
  {G{\'o}mez-Santos}(2012)}]{Stauber:2012aa}%
  \BibitemOpen
  \bibfield  {author} {\bibinfo {author} {\bibfnamefont {T.}~\bibnamefont
  {Stauber}}\ and\ \bibinfo {author} {\bibfnamefont {G.}~\bibnamefont
  {G{\'o}mez-Santos}},\ }\bibfield  {booktitle} {\emph {\bibinfo {booktitle}
  {New Journal of Physics}},\ }\href {\doibase 10.1088/1367-2630/14/10/105018}
  {\bibfield  {journal} {\bibinfo  {journal} {New Journal of Physics}\ }\textbf
  {\bibinfo {volume} {14}},\ \bibinfo {pages} {105018} (\bibinfo {year}
  {2012})}\BibitemShut {NoStop}%
\bibitem [{\citenamefont {Rodrigo}\ \emph {et~al.}(2017)\citenamefont
  {Rodrigo}, \citenamefont {Tittl}, \citenamefont {Limaj}, \citenamefont
  {Abajo}, \citenamefont {Pruneri},\ and\ \citenamefont
  {Altug}}]{Rodrigo:2017aa}%
  \BibitemOpen
  \bibfield  {author} {\bibinfo {author} {\bibfnamefont {D.}~\bibnamefont
  {Rodrigo}}, \bibinfo {author} {\bibfnamefont {A.}~\bibnamefont {Tittl}},
  \bibinfo {author} {\bibfnamefont {O.}~\bibnamefont {Limaj}}, \bibinfo
  {author} {\bibfnamefont {F.~J. G.~d.}\ \bibnamefont {Abajo}}, \bibinfo
  {author} {\bibfnamefont {V.}~\bibnamefont {Pruneri}}, \ and\ \bibinfo
  {author} {\bibfnamefont {H.}~\bibnamefont {Altug}},\ }\href {\doibase
  10.1038/lsa.2016.277} {\bibfield  {journal} {\bibinfo  {journal} {Light:
  Science \& Applications}\ }\textbf {\bibinfo {volume} {6}},\ \bibinfo {pages}
  {e16277} (\bibinfo {year} {2017})}\BibitemShut {NoStop}%
\bibitem [{\citenamefont {Das~Sarma}\ and\ \citenamefont
  {Madhukar}(1981)}]{Das-Sarma:1981aa}%
  \BibitemOpen
  \bibfield  {author} {\bibinfo {author} {\bibfnamefont {S.}~\bibnamefont
  {Das~Sarma}}\ and\ \bibinfo {author} {\bibfnamefont {A.}~\bibnamefont
  {Madhukar}},\ }\href {\doibase 10.1103/PhysRevB.23.805} {\bibfield  {journal}
  {\bibinfo  {journal} {Physical Review B}\ }\textbf {\bibinfo {volume} {23}},\
  \bibinfo {pages} {805} (\bibinfo {year} {1981})}\BibitemShut {NoStop}%
\bibitem [{\citenamefont {Duan}\ \emph {et~al.}(2000)\citenamefont {Duan},
  \citenamefont {Giedke}, \citenamefont {Cirac},\ and\ \citenamefont
  {Zoller}}]{Duan_2000}%
  \BibitemOpen
  \bibfield  {author} {\bibinfo {author} {\bibfnamefont {L.-M.}\ \bibnamefont
  {Duan}}, \bibinfo {author} {\bibfnamefont {G.}~\bibnamefont {Giedke}},
  \bibinfo {author} {\bibfnamefont {J.~I.}\ \bibnamefont {Cirac}}, \ and\
  \bibinfo {author} {\bibfnamefont {P.}~\bibnamefont {Zoller}},\ }\href
  {\doibase 10.1103/PhysRevLett.84.2722} {\bibfield  {journal} {\bibinfo
  {journal} {Phys. Rev. Lett.}\ }\textbf {\bibinfo {volume} {84}},\ \bibinfo
  {pages} {2722} (\bibinfo {year} {2000})}\BibitemShut {NoStop}%
\bibitem [{\citenamefont {Braunstein}\ and\ \citenamefont
  {Pati}(2003)}]{Braunstein-book}%
  \BibitemOpen
  \bibfield  {author} {\bibinfo {author} {\bibfnamefont {S.}~\bibnamefont
  {Braunstein}}\ and\ \bibinfo {author} {\bibfnamefont {A.~K.}\ \bibnamefont
  {Pati}},\ }\href@noop {} {\emph {\bibinfo {title} {Quantum Information with
  Continuous Variables}}}\ (\bibinfo  {publisher} {Kluwer Academic
  Publishers},\ \bibinfo {year} {2003})\BibitemShut {NoStop}%
\bibitem [{\citenamefont {Wunsch}\ \emph {et~al.}(2006)\citenamefont {Wunsch},
  \citenamefont {Stauber}, \citenamefont {Sols},\ and\ \citenamefont
  {Guinea}}]{Wunsch:2006aa}%
  \BibitemOpen
  \bibfield  {author} {\bibinfo {author} {\bibfnamefont {B.}~\bibnamefont
  {Wunsch}}, \bibinfo {author} {\bibfnamefont {T.}~\bibnamefont {Stauber}},
  \bibinfo {author} {\bibfnamefont {F.}~\bibnamefont {Sols}}, \ and\ \bibinfo
  {author} {\bibfnamefont {F.}~\bibnamefont {Guinea}},\ }\href {\doibase
  10.1088/1367-2630/8/12/318} {\bibfield  {journal} {\bibinfo  {journal} {New
  Journal of Physics}\ }\textbf {\bibinfo {volume} {8}},\ \bibinfo {pages}
  {318} (\bibinfo {year} {2006})}\BibitemShut {NoStop}%
\bibitem [{\citenamefont {Hwang}\ and\ \citenamefont
  {Das~Sarma}(2007)}]{Hwang:2007aa}%
  \BibitemOpen
  \bibfield  {author} {\bibinfo {author} {\bibfnamefont {E.~H.}\ \bibnamefont
  {Hwang}}\ and\ \bibinfo {author} {\bibfnamefont {S.}~\bibnamefont
  {Das~Sarma}},\ }\href {\doibase 10.1103/PhysRevB.75.205418} {\bibfield
  {journal} {\bibinfo  {journal} {Physical Review B}\ }\textbf {\bibinfo
  {volume} {75}},\ \bibinfo {pages} {205418} (\bibinfo {year}
  {2007})}\BibitemShut {NoStop}%
\bibitem [{\citenamefont {Brey}\ and\ \citenamefont
  {Fertig}(2007)}]{Brey:2007aa}%
  \BibitemOpen
  \bibfield  {author} {\bibinfo {author} {\bibfnamefont {L.}~\bibnamefont
  {Brey}}\ and\ \bibinfo {author} {\bibfnamefont {H.~A.}\ \bibnamefont
  {Fertig}},\ }\href {\doibase 10.1103/PhysRevB.75.125434} {\bibfield
  {journal} {\bibinfo  {journal} {Physical Review B}\ }\textbf {\bibinfo
  {volume} {75}},\ \bibinfo {pages} {125434} (\bibinfo {year}
  {2007})}\BibitemShut {NoStop}%
\bibitem [{Sup()}]{Supp1}%
  \BibitemOpen
  \href@noop {} {}\bibinfo {note} {See Supplemental Material for more details
  with additional analytical results.}\BibitemShut {Stop}%
\bibitem [{\citenamefont {Elson}\ and\ \citenamefont
  {Ritchie}(1971)}]{Elson:1971aa}%
  \BibitemOpen
  \bibfield  {author} {\bibinfo {author} {\bibfnamefont {J.~M.}\ \bibnamefont
  {Elson}}\ and\ \bibinfo {author} {\bibfnamefont {R.~H.}\ \bibnamefont
  {Ritchie}},\ }\href {\doibase 10.1103/PhysRevB.4.4129} {\bibfield  {journal}
  {\bibinfo  {journal} {Physical Review B}\ }\textbf {\bibinfo {volume} {4}},\
  \bibinfo {pages} {4129} (\bibinfo {year} {1971})}\BibitemShut {NoStop}%
\bibitem [{\citenamefont {Gruner}\ and\ \citenamefont
  {Welsch}(1996)}]{Gruner:1996aa}%
  \BibitemOpen
  \bibfield  {author} {\bibinfo {author} {\bibfnamefont {T.}~\bibnamefont
  {Gruner}}\ and\ \bibinfo {author} {\bibfnamefont {D.~G.}\ \bibnamefont
  {Welsch}},\ }\href {\doibase 10.1103/PhysRevA.53.1818} {\bibfield  {journal}
  {\bibinfo  {journal} {Physical Review A}\ }\textbf {\bibinfo {volume} {53}},\
  \bibinfo {pages} {1818} (\bibinfo {year} {1996})}\BibitemShut {NoStop}%
\bibitem [{\citenamefont {Archambault}\ \emph {et~al.}(2010)\citenamefont
  {Archambault}, \citenamefont {Marquier}, \citenamefont {Greffet},\ and\
  \citenamefont {Arnold}}]{Archambault:2010aa}%
  \BibitemOpen
  \bibfield  {author} {\bibinfo {author} {\bibfnamefont {A.}~\bibnamefont
  {Archambault}}, \bibinfo {author} {\bibfnamefont {F.}~\bibnamefont
  {Marquier}}, \bibinfo {author} {\bibfnamefont {J.-J.}\ \bibnamefont
  {Greffet}}, \ and\ \bibinfo {author} {\bibfnamefont {C.}~\bibnamefont
  {Arnold}},\ }\href {\doibase 10.1103/PhysRevB.82.035411} {\bibfield
  {journal} {\bibinfo  {journal} {Physical Review B}\ }\textbf {\bibinfo
  {volume} {82}},\ \bibinfo {pages} {035411} (\bibinfo {year}
  {2010})}\BibitemShut {NoStop}%
\bibitem [{\citenamefont {Hanson}\ \emph {et~al.}(2015)\citenamefont {Hanson},
  \citenamefont {Hassani~Gangaraj}, \citenamefont {Lee}, \citenamefont
  {Angelakis},\ and\ \citenamefont {Tame}}]{Hanson:2015aa}%
  \BibitemOpen
  \bibfield  {author} {\bibinfo {author} {\bibfnamefont {G.~W.}\ \bibnamefont
  {Hanson}}, \bibinfo {author} {\bibfnamefont {S.~A.}\ \bibnamefont
  {Hassani~Gangaraj}}, \bibinfo {author} {\bibfnamefont {C.}~\bibnamefont
  {Lee}}, \bibinfo {author} {\bibfnamefont {D.~G.}\ \bibnamefont {Angelakis}},
  \ and\ \bibinfo {author} {\bibfnamefont {M.}~\bibnamefont {Tame}},\ }\href
  {\doibase 10.1103/PhysRevA.92.013828} {\bibfield  {journal} {\bibinfo
  {journal} {Physical Review A}\ }\textbf {\bibinfo {volume} {92}},\ \bibinfo
  {pages} {013828} (\bibinfo {year} {2015})}\BibitemShut {NoStop}%
\bibitem [{\citenamefont {Ferreira}\ \emph {et~al.}(2020)\citenamefont
  {Ferreira}, \citenamefont {Amorim}, \citenamefont {Chaves},\ and\
  \citenamefont {Peres}}]{Ferreira:2020aa}%
  \BibitemOpen
  \bibfield  {author} {\bibinfo {author} {\bibfnamefont {B.~A.}\ \bibnamefont
  {Ferreira}}, \bibinfo {author} {\bibfnamefont {B.}~\bibnamefont {Amorim}},
  \bibinfo {author} {\bibfnamefont {A.~J.}\ \bibnamefont {Chaves}}, \ and\
  \bibinfo {author} {\bibfnamefont {N.~M.~R.}\ \bibnamefont {Peres}},\ }\href
  {\doibase 10.1103/PhysRevA.101.033817} {\bibfield  {journal} {\bibinfo
  {journal} {Physical Review A}\ }\textbf {\bibinfo {volume} {101}},\ \bibinfo
  {pages} {033817} (\bibinfo {year} {2020})}\BibitemShut {NoStop}%
\bibitem [{\citenamefont {Hwang}\ and\ \citenamefont
  {Das~Sarma}(2009)}]{Hwang:2009aa}%
  \BibitemOpen
  \bibfield  {author} {\bibinfo {author} {\bibfnamefont {E.~H.}\ \bibnamefont
  {Hwang}}\ and\ \bibinfo {author} {\bibfnamefont {S.}~\bibnamefont
  {Das~Sarma}},\ }\href {\doibase 10.1103/PhysRevB.80.205405} {\bibfield
  {journal} {\bibinfo  {journal} {Physical Review B}\ }\textbf {\bibinfo
  {volume} {80}},\ \bibinfo {pages} {205405} (\bibinfo {year}
  {2009})}\BibitemShut {NoStop}%
\bibitem [{\citenamefont {Alexey V.~Kavokin}\ and\ \citenamefont
  {Laussy}(2007)}]{Kavokin-Book}%
  \BibitemOpen
  \bibfield  {author} {\bibinfo {author} {\bibfnamefont {G.~M.}\ \bibnamefont
  {Alexey V.~Kavokin}, \bibfnamefont {Jeremy J.~Baumberg}}\ and\ \bibinfo
  {author} {\bibfnamefont {F.~P.}\ \bibnamefont {Laussy}},\ }\href@noop {}
  {\emph {\bibinfo {title} {Microcavities}}}\ (\bibinfo  {publisher} {Oxford
  University Press},\ \bibinfo {year} {2007})\BibitemShut {NoStop}%
\bibitem [{\citenamefont {Frisk~Kockum}\ \emph {et~al.}(2019)\citenamefont
  {Frisk~Kockum}, \citenamefont {Miranowicz}, \citenamefont {De~Liberato},
  \citenamefont {Savasta},\ and\ \citenamefont {Nori}}]{Frisk-Kockum:2019aa}%
  \BibitemOpen
  \bibfield  {author} {\bibinfo {author} {\bibfnamefont {A.}~\bibnamefont
  {Frisk~Kockum}}, \bibinfo {author} {\bibfnamefont {A.}~\bibnamefont
  {Miranowicz}}, \bibinfo {author} {\bibfnamefont {S.}~\bibnamefont
  {De~Liberato}}, \bibinfo {author} {\bibfnamefont {S.}~\bibnamefont
  {Savasta}}, \ and\ \bibinfo {author} {\bibfnamefont {F.}~\bibnamefont
  {Nori}},\ }\href {\doibase 10.1038/s42254-018-0006-2} {\bibfield  {journal}
  {\bibinfo  {journal} {Nature Reviews Physics}\ }\textbf {\bibinfo {volume}
  {1}},\ \bibinfo {pages} {19} (\bibinfo {year} {2019})}\BibitemShut {NoStop}%
\bibitem [{\citenamefont {Kirton}\ \emph {et~al.}(2019)\citenamefont {Kirton},
  \citenamefont {Roses}, \citenamefont {Keeling},\ and\ \citenamefont
  {Dalla~Torre}}]{Kirton:2019aa}%
  \BibitemOpen
  \bibfield  {author} {\bibinfo {author} {\bibfnamefont {P.}~\bibnamefont
  {Kirton}}, \bibinfo {author} {\bibfnamefont {M.~M.}\ \bibnamefont {Roses}},
  \bibinfo {author} {\bibfnamefont {J.}~\bibnamefont {Keeling}}, \ and\
  \bibinfo {author} {\bibfnamefont {E.~G.}\ \bibnamefont {Dalla~Torre}},\
  }\bibfield  {booktitle} {\emph {\bibinfo {booktitle} {Advanced Quantum
  Technologies}},\ }\href {\doibase https://doi.org/10.1002/qute.201800043}
  {\bibfield  {journal} {\bibinfo  {journal} {Advanced Quantum Technologies}\
  }\textbf {\bibinfo {volume} {2}},\ \bibinfo {pages} {1800043} (\bibinfo
  {year} {2019})}\BibitemShut {NoStop}%
\bibitem [{\citenamefont {Hopfield}(1958)}]{Hopfield:1958aa}%
  \BibitemOpen
  \bibfield  {author} {\bibinfo {author} {\bibfnamefont {J.~J.}\ \bibnamefont
  {Hopfield}},\ }\href {\doibase 10.1103/PhysRev.112.1555} {\bibfield
  {journal} {\bibinfo  {journal} {Physical Review}\ }\textbf {\bibinfo {volume}
  {112}},\ \bibinfo {pages} {1555} (\bibinfo {year} {1958})}\BibitemShut
  {NoStop}%
\bibitem [{\citenamefont {Makarov}(2018)}]{Makarov}%
  \BibitemOpen
  \bibfield  {author} {\bibinfo {author} {\bibfnamefont {D.~N.}\ \bibnamefont
  {Makarov}},\ }\href@noop {} {\bibfield  {journal} {\bibinfo  {journal}
  {Physical Review E}\ }\textbf {\bibinfo {volume} {97}},\ \bibinfo {pages}
  {042203} (\bibinfo {year} {2018})}\BibitemShut {NoStop}%
\bibitem [{\citenamefont {Zhou}\ \emph {et~al.}(2020)\citenamefont {Zhou},
  \citenamefont {Zhou}, \citenamefont {Yin}, \citenamefont {Huang},\ and\
  \citenamefont {Liao}}]{Zhou_2020}%
  \BibitemOpen
  \bibfield  {author} {\bibinfo {author} {\bibfnamefont {J.-Y.}\ \bibnamefont
  {Zhou}}, \bibinfo {author} {\bibfnamefont {Y.-H.}\ \bibnamefont {Zhou}},
  \bibinfo {author} {\bibfnamefont {X.-L.}\ \bibnamefont {Yin}}, \bibinfo
  {author} {\bibfnamefont {J.-F.}\ \bibnamefont {Huang}}, \ and\ \bibinfo
  {author} {\bibfnamefont {J.-Q.}\ \bibnamefont {Liao}},\ }\href@noop {}
  {\bibfield  {journal} {\bibinfo  {journal} {Scientific Reports}\ }\textbf
  {\bibinfo {volume} {10}},\ \bibinfo {pages} {12557} (\bibinfo {year}
  {2020})}\BibitemShut {NoStop}%
\bibitem [{\citenamefont {Girvin}\ and\ \citenamefont
  {Yang}(2019)}]{Girvin-book}%
  \BibitemOpen
  \bibfield  {author} {\bibinfo {author} {\bibfnamefont {S.~M.}\ \bibnamefont
  {Girvin}}\ and\ \bibinfo {author} {\bibfnamefont {K.}~\bibnamefont {Yang}},\
  }\href@noop {} {\emph {\bibinfo {title} {Modern Condensed Matter Physics}}}\
  (\bibinfo  {publisher} {Cambridge University Press},\ \bibinfo {address}
  {Cambridge},\ \bibinfo {year} {2019})\BibitemShut {NoStop}%
\bibitem [{\citenamefont {Nation}\ \emph {et~al.}(2012)\citenamefont {Nation},
  \citenamefont {Johansson}, \citenamefont {Blencowe},\ and\ \citenamefont
  {Nori}}]{Nation_2012}%
  \BibitemOpen
  \bibfield  {author} {\bibinfo {author} {\bibfnamefont {P.~D.}\ \bibnamefont
  {Nation}}, \bibinfo {author} {\bibfnamefont {J.~R.}\ \bibnamefont
  {Johansson}}, \bibinfo {author} {\bibfnamefont {M.~P.}\ \bibnamefont
  {Blencowe}}, \ and\ \bibinfo {author} {\bibfnamefont {F.}~\bibnamefont
  {Nori}},\ }\href {\doibase 10.1103/RevModPhys.84.1} {\bibfield  {journal}
  {\bibinfo  {journal} {Rev. Mod. Phys.}\ }\textbf {\bibinfo {volume} {84}},\
  \bibinfo {pages} {1} (\bibinfo {year} {2012})}\BibitemShut {NoStop}%
\bibitem [{\citenamefont {Leonhardt}(2010)}]{Leonhardt-book}%
  \BibitemOpen
  \bibfield  {author} {\bibinfo {author} {\bibfnamefont {U.}~\bibnamefont
  {Leonhardt}},\ }\href@noop {} {\emph {\bibinfo {title} {Essential Quantum
  Optics}}}\ (\bibinfo  {publisher} {Cambridge University Press},\ \bibinfo
  {year} {2010})\BibitemShut {NoStop}%
\bibitem [{\citenamefont {Chen}\ \emph
  {et~al.}(2012{\natexlab{b}})\citenamefont {Chen}, \citenamefont {Badioli},
  \citenamefont {Alonso-Gonzalez}, \citenamefont {Thongrattanasiri},
  \citenamefont {Huth}, \citenamefont {Osmond}, \citenamefont {Spasenovic},
  \citenamefont {Centeno}, \citenamefont {Pesquera}, \citenamefont {Godignon},
  \citenamefont {Zurutuza~Elorza}, \citenamefont {Camara}, \citenamefont
  {de~Abajo}, \citenamefont {Hillenbrand},\ and\ \citenamefont
  {Koppens}}]{Chen12}%
  \BibitemOpen
  \bibfield  {author} {\bibinfo {author} {\bibfnamefont {J.}~\bibnamefont
  {Chen}}, \bibinfo {author} {\bibfnamefont {M.}~\bibnamefont {Badioli}},
  \bibinfo {author} {\bibfnamefont {P.}~\bibnamefont {Alonso-Gonzalez}},
  \bibinfo {author} {\bibfnamefont {S.}~\bibnamefont {Thongrattanasiri}},
  \bibinfo {author} {\bibfnamefont {F.}~\bibnamefont {Huth}}, \bibinfo {author}
  {\bibfnamefont {J.}~\bibnamefont {Osmond}}, \bibinfo {author} {\bibfnamefont
  {M.}~\bibnamefont {Spasenovic}}, \bibinfo {author} {\bibfnamefont
  {A.}~\bibnamefont {Centeno}}, \bibinfo {author} {\bibfnamefont
  {A.}~\bibnamefont {Pesquera}}, \bibinfo {author} {\bibfnamefont
  {P.}~\bibnamefont {Godignon}}, \bibinfo {author} {\bibfnamefont
  {A.}~\bibnamefont {Zurutuza~Elorza}}, \bibinfo {author} {\bibfnamefont
  {N.}~\bibnamefont {Camara}}, \bibinfo {author} {\bibfnamefont {F.~J.~G.}\
  \bibnamefont {de~Abajo}}, \bibinfo {author} {\bibfnamefont {R.}~\bibnamefont
  {Hillenbrand}}, \ and\ \bibinfo {author} {\bibfnamefont {F.~H.~L.}\
  \bibnamefont {Koppens}},\ }\href {\doibase 10.1038/nature11254} {\bibfield
  {journal} {\bibinfo  {journal} {Nature}\ }\textbf {\bibinfo {volume} {487}},\
  \bibinfo {pages} {77} (\bibinfo {year} {2012}{\natexlab{b}})}\BibitemShut
  {NoStop}%
\bibitem [{\citenamefont {Fei}\ \emph {et~al.}(2012{\natexlab{b}})\citenamefont
  {Fei}, \citenamefont {Rodin}, \citenamefont {Andreev}, \citenamefont {Bao},
  \citenamefont {McLeod}, \citenamefont {Wagner}, \citenamefont {Zhang},
  \citenamefont {Zhao}, \citenamefont {Thiemens}, \citenamefont {Dominguez},
  \citenamefont {Fogler}, \citenamefont {Neto}, \citenamefont {Lau},
  \citenamefont {Keilmann},\ and\ \citenamefont {Basov}}]{Fei12}%
  \BibitemOpen
  \bibfield  {author} {\bibinfo {author} {\bibfnamefont {Z.}~\bibnamefont
  {Fei}}, \bibinfo {author} {\bibfnamefont {A.~S.}\ \bibnamefont {Rodin}},
  \bibinfo {author} {\bibfnamefont {G.~O.}\ \bibnamefont {Andreev}}, \bibinfo
  {author} {\bibfnamefont {W.}~\bibnamefont {Bao}}, \bibinfo {author}
  {\bibfnamefont {A.~S.}\ \bibnamefont {McLeod}}, \bibinfo {author}
  {\bibfnamefont {M.}~\bibnamefont {Wagner}}, \bibinfo {author} {\bibfnamefont
  {L.~M.}\ \bibnamefont {Zhang}}, \bibinfo {author} {\bibfnamefont
  {Z.}~\bibnamefont {Zhao}}, \bibinfo {author} {\bibfnamefont {M.}~\bibnamefont
  {Thiemens}}, \bibinfo {author} {\bibfnamefont {G.}~\bibnamefont {Dominguez}},
  \bibinfo {author} {\bibfnamefont {M.~M.}\ \bibnamefont {Fogler}}, \bibinfo
  {author} {\bibfnamefont {A.~H.~C.}\ \bibnamefont {Neto}}, \bibinfo {author}
  {\bibfnamefont {C.~N.}\ \bibnamefont {Lau}}, \bibinfo {author} {\bibfnamefont
  {F.}~\bibnamefont {Keilmann}}, \ and\ \bibinfo {author} {\bibfnamefont
  {D.~N.}\ \bibnamefont {Basov}},\ }\href {\doibase 10.1038/nature11253}
  {\bibfield  {journal} {\bibinfo  {journal} {Nature}\ }\textbf {\bibinfo
  {volume} {487}},\ \bibinfo {pages} {82} (\bibinfo {year}
  {2012}{\natexlab{b}})}\BibitemShut {NoStop}%
\bibitem [{\citenamefont {Woessner}\ \emph {et~al.}(2015)\citenamefont
  {Woessner}, \citenamefont {Lundeberg}, \citenamefont {Gao}, \citenamefont
  {Principi}, \citenamefont {Alonso-Gonz{\'a}lez}, \citenamefont {Carrega},
  \citenamefont {Watanabe}, \citenamefont {Taniguchi}, \citenamefont {Vignale},
  \citenamefont {Polini}, \citenamefont {Hone}, \citenamefont {Hillenbrand},\
  and\ \citenamefont {Koppens}}]{Woessner:2015aa}%
  \BibitemOpen
  \bibfield  {author} {\bibinfo {author} {\bibfnamefont {A.}~\bibnamefont
  {Woessner}}, \bibinfo {author} {\bibfnamefont {M.~B.}\ \bibnamefont
  {Lundeberg}}, \bibinfo {author} {\bibfnamefont {Y.}~\bibnamefont {Gao}},
  \bibinfo {author} {\bibfnamefont {A.}~\bibnamefont {Principi}}, \bibinfo
  {author} {\bibfnamefont {P.}~\bibnamefont {Alonso-Gonz{\'a}lez}}, \bibinfo
  {author} {\bibfnamefont {M.}~\bibnamefont {Carrega}}, \bibinfo {author}
  {\bibfnamefont {K.}~\bibnamefont {Watanabe}}, \bibinfo {author}
  {\bibfnamefont {T.}~\bibnamefont {Taniguchi}}, \bibinfo {author}
  {\bibfnamefont {G.}~\bibnamefont {Vignale}}, \bibinfo {author} {\bibfnamefont
  {M.}~\bibnamefont {Polini}}, \bibinfo {author} {\bibfnamefont
  {J.}~\bibnamefont {Hone}}, \bibinfo {author} {\bibfnamefont {R.}~\bibnamefont
  {Hillenbrand}}, \ and\ \bibinfo {author} {\bibfnamefont {F.~H.~L.}\
  \bibnamefont {Koppens}},\ }\href {\doibase 10.1038/nmat4169} {\bibfield
  {journal} {\bibinfo  {journal} {Nature Materials}\ }\textbf {\bibinfo
  {volume} {14}},\ \bibinfo {pages} {421} (\bibinfo {year} {2015})}\BibitemShut
  {NoStop}%
\bibitem [{Mai()}]{Main1}%
  \BibitemOpen
  \href@noop {} {}\bibinfo {note} {For example, for plasmons of sufficiently
  high frequency to be easily detected (bigger than 1THz), for ultra strong
  coupling the electron density would have to vanish at roughly this rate or
  faster. Fully depleting an electron layer on a time scale, of the order of
  picoseconds, for most two-dimensional materials is difficult.}\BibitemShut
  {Stop}%
\bibitem [{\citenamefont {Tsui}\ \emph {et~al.}(1980)\citenamefont {Tsui},
  \citenamefont {Gornik},\ and\ \citenamefont {Logan}}]{Tsui:1980aa}%
  \BibitemOpen
  \bibfield  {author} {\bibinfo {author} {\bibfnamefont {D.~C.}\ \bibnamefont
  {Tsui}}, \bibinfo {author} {\bibfnamefont {E.}~\bibnamefont {Gornik}}, \ and\
  \bibinfo {author} {\bibfnamefont {R.~A.}\ \bibnamefont {Logan}},\ }\href
  {\doibase https://doi.org/10.1016/0038-1098(80)91043-1} {\bibfield  {journal}
  {\bibinfo  {journal} {Solid State Communications}\ }\textbf {\bibinfo
  {volume} {35}},\ \bibinfo {pages} {875} (\bibinfo {year} {1980})}\BibitemShut
  {NoStop}%
\bibitem [{\citenamefont {H{\"o}pfel}\ \emph {et~al.}(1982)\citenamefont
  {H{\"o}pfel}, \citenamefont {Vass},\ and\ \citenamefont
  {Gornik}}]{Hopfel:1982aa}%
  \BibitemOpen
  \bibfield  {author} {\bibinfo {author} {\bibfnamefont {R.~A.}\ \bibnamefont
  {H{\"o}pfel}}, \bibinfo {author} {\bibfnamefont {E.}~\bibnamefont {Vass}}, \
  and\ \bibinfo {author} {\bibfnamefont {E.}~\bibnamefont {Gornik}},\ }\href
  {\doibase 10.1103/PhysRevLett.49.1667} {\bibfield  {journal} {\bibinfo
  {journal} {Physical Review Letters}\ }\textbf {\bibinfo {volume} {49}},\
  \bibinfo {pages} {1667} (\bibinfo {year} {1982})}\BibitemShut {NoStop}%
\bibitem [{\citenamefont {Okisu}\ \emph {et~al.}(1986)\citenamefont {Okisu},
  \citenamefont {Sambe},\ and\ \citenamefont {Kobayashi}}]{Okisu:1986aa}%
  \BibitemOpen
  \bibfield  {author} {\bibinfo {author} {\bibfnamefont {N.}~\bibnamefont
  {Okisu}}, \bibinfo {author} {\bibfnamefont {Y.}~\bibnamefont {Sambe}}, \ and\
  \bibinfo {author} {\bibfnamefont {T.}~\bibnamefont {Kobayashi}},\ }\href
  {\doibase 10.1063/1.96718} {\bibfield  {journal} {\bibinfo  {journal}
  {Applied Physics Letters}\ }\textbf {\bibinfo {volume} {48}},\ \bibinfo
  {pages} {776} (\bibinfo {year} {1986})}\BibitemShut {NoStop}%
\bibitem [{\citenamefont {Li}\ \emph {et~al.}(2019)\citenamefont {Li},
  \citenamefont {Ferreyra}, \citenamefont {Swan},\ and\ \citenamefont
  {Paiella}}]{Li:2019aa}%
  \BibitemOpen
  \bibfield  {author} {\bibinfo {author} {\bibfnamefont {Y.}~\bibnamefont
  {Li}}, \bibinfo {author} {\bibfnamefont {P.}~\bibnamefont {Ferreyra}},
  \bibinfo {author} {\bibfnamefont {A.~K.}\ \bibnamefont {Swan}}, \ and\
  \bibinfo {author} {\bibfnamefont {R.}~\bibnamefont {Paiella}},\ }\bibfield
  {booktitle} {\emph {\bibinfo {booktitle} {ACS Photonics}},\ }\href {\doibase
  10.1021/acsphotonics.9b01037} {\bibfield  {journal} {\bibinfo  {journal} {ACS
  Photonics}\ }\textbf {\bibinfo {volume} {6}},\ \bibinfo {pages} {2562}
  (\bibinfo {year} {2019})}\BibitemShut {NoStop}%
\end{thebibliography}
\vspace{0.5truecm}
\end{document}